\documentclass[prx,aps,superscriptaddress,footinbib,twocolumn]{revtex4-1}

\usepackage{mathrsfs}
\usepackage{amsfonts}
\usepackage{amssymb}
\usepackage{amsmath}
\usepackage{graphicx}
\usepackage[usenames,dvipsnames]{color}
\usepackage[colorlinks=true,citecolor=blue,linkcolor=magenta]{hyperref}
\usepackage{ulem}
\usepackage{lmodern}
\usepackage{amsthm}
\usepackage{dsfont}
\usepackage{natbib}

\usepackage[noend]{algpseudocode}
\usepackage{algorithm}
\newcommand{\figpath}{.}

\renewcommand{\Re}{\mathrm{Re}}
\renewcommand{\Im}{\mathrm{Im}}
\newcommand{\Tr}{\mathrm{Tr}}

\newcommand{\abs}[1]{\vert #1 \vert}
\newcommand{\absLR}[1]{\left\vert #1 \right\vert}

\newcommand{\ket}[1]{\vert{ #1 }\rangle}
\newcommand{\bra}[1]{\langle{ #1 }\vert}
\newcommand{\ketbra}[2]{\vert #1 \rangle \langle #2 \vert}
\newcommand{\braket}[2]{\langle #1 \vert #2 \rangle}
\newcommand{\mean}[1]{\langle #1 \rangle}
\newcommand{\meanLR}[1]{\left\langle #1 \right\rangle}

\newcommand{\st}{\,\vert\,}

\newcommand{\calG}{\mathcal{G}}

\newcommand{\bfr}{\boldsymbol{r}}
\newcommand{\bfs}{\boldsymbol{s}}
\newcommand{\bfP}{\boldsymbol{P}}
\newcommand{\bfh}{\boldsymbol{h}}
\newcommand{\bfsigma}{\boldsymbol{\sigma}}

\newcommand{\bfx}{\boldsymbol{x}}
\newcommand{\bfC}{\boldsymbol{C}}

\begin{document}

\title{Accelerated quantum Monte Carlo with mitigated error on noisy quantum computer}

\author{Yongdan Yang}
\affiliation{Graduate School of China Academy of Engineering Physics, Beijing 100193, China}

\author{Bing-Nan Lu}
\affiliation{Graduate School of China Academy of Engineering Physics, Beijing 100193, China}

\author{Ying Li}
\email{yli@gscaep.ac.cn}
\affiliation{Graduate School of China Academy of Engineering Physics, Beijing 100193, China}

\begin{abstract}
Quantum Monte Carlo and quantum simulation are both important tools for understanding quantum many-body systems. As a classical algorithm, quantum Monte Carlo suffers from the sign problem, preventing its application to most fermion systems and real time dynamics. In this paper, we introduce a novel non-variational algorithm using quantum simulation as a subroutine to accelerate quantum Monte Carlo by easing the sign problem. The quantum subroutine can be implemented with shallow circuits and, by incorporating error mitigation, can reduce the Monte Carlo variance by several orders of magnitude even when the circuit noise is significant. As such, the proposed quantum algorithm is applicable to near-term noisy quantum hardware.
\end{abstract}

\maketitle

\section{Introduction}

The simulation of quantum many-body systems is one of the main motivations for quantum computing~\cite{Feynman1982}. A lot of quantum many-body problems are intractable in classical computing. An apparent reason is that the Hilbert space dimension increases exponentially with the system size and it is impossible to store the wave function of a large system in classical memory. Quantum Monte Carlo (QMC) is a group of classical algorithms designed to bypass this memory issue. By sampling only the most important part of the configuration space, QMC can solve certain many-body problems at a polynomial complexity, at the cost of introducing small statistical errors. Unfortunately, when applied to fermion systems and real time dynamics, QMC encounters the notorious sign problem, i.e.~the target amplitude is a highly-oscillating function with alternating sign. This sign problem results in a variance that increases exponentially in the Monte Carlo simulation~\cite{Troyer2005}, forming the dominant limitation of QMC. On the other hand, by mapping the target wave function of the simulated system into the wave function of qubits on a fault-tolerant quantum computer~\cite{Knill1998}, we can reproduce the dynamics of quantum systems while the memory and run time scale polynomially~\cite{Lloyd1996}. With the development of the fault-tolerant technologies as a long-term goal, exploring the power of noisy intermediate-scale quantum hardware is of particular importance for near-term applications~\cite{Preskill2018}. In this paper, we establish the framework of quantum-circuit Monte Carlo (QCMC) algorithm, in which quantum computing is a subroutine of QMC. We show that this algorithm has a quantum advantage in solving many-body problems, even on noisy quantum computers.

Since Ulam and Metropolis's pioneering work of using random sampling to simulate real physical systems~\cite{Metropolis1949}, the Monte Carlo method has grown into a large family of algorithms. Here, we focus on a specific subset of Monte Carlo algorithms, namely, the QMC methods, which are based on real or imaginary time evolution. These methods include Green's function Monte Carlo~\cite{Carlson2015}, auxiliary field Monte Carlo~\cite{Blankenbecler1981, Lee2009}, world-line Monte Carlo~\cite{Evertz1993, Bour2015}, and diagrammatic Monte Carlo~\cite{Houcke2010, Houcke2012, Cohen2015, Bertrand2019}, and their various variants. In what follows, by QMC, we refer to this subset of algorithms. The other QMC algorithms are based on variational methods~\cite{Lomnitz1981} but while their connection to quantum computing is also an interesting topic, they are not be covered in this work.

In most QMC methods, we sample the configurations according to a quasi-probability amplitude derived from time evolution. For fermion systems such an amplitude is usually a complex number, which can be positive definite if the system respects certain symmetries. Examples of the latter case include the half-filled Hubbard model with particle-hole exchange symmetry~\cite{Hubbard1963, Takahashi1977} and the nuclear system with Wigner-SU(4) symmetry~\cite{Lu2019, Lee2020}. However, a realistic Hamiltonian usually contains terms that break these symmetries and induce oscillating phases in the probability amplitude. As a result, even though QMC methods are very successful in describing certain strongly correlated systems in chemistry~\cite{Hammond1994}, condensed matter physics~\cite{Foulkes2001}, and nuclear physics~\cite{Carlson2015}, their application is still rather limited due to the sign problem. Although in some important cases the sign problem can be alleviated using complicated techniques~\cite{Hangleiter2020}, e.g.~the complex Langevin method~\cite{Parisi1983, Klauder1983} or the Lefschetz thimble method~\cite{Cristoforetti2012, Wynen2021}, finding a generic solution is unlikely, as it is proven that the sign problem is NP-hard~\cite{Troyer2005}.

In quantum computing, the qubit and time costs for simulating the unitary time evolution of a quantum system scale polynomially with the problem parameters, i.e.~the system size, evolution time, and accuracy. Such algorithms include the Lie-Trotter-Suzuki decomposition~\cite{Lloyd1996, Berry2007, Wiebe2010}, the truncated Taylor series~\cite{Berry2015, Meister2020}, linear combinations of Lie-Trotter-Suzuki products~\cite{Childs2012, Faehrmann2021}, and the random compiler~\cite{Campbell2019}. Based on the simulation of unitary time evolution, one can also simulate open-system dynamics~\cite{Kliesch2011, Wang2011}, solve equilibrium-state problems~\cite{Temme2011, Riera2012} and find the ground state for certain Hamiltonians~\cite{OBrien2019, Lu2021, Turro2021}. However, implementation of these algorithms at a meaningful scale usually requires a fault-tolerant quantum computer~\cite{Reiher2017, Babbush2018}, on which the logical error rate can be reduced to any level at a polynomial cost in quantum error correction~\cite{Fowler2012}. In recent years, hybrid quantum-classical algorithms have been developed for applications before the era of fault-tolerant technologies~\cite{Bauer2016}. Many such algorithms are based on variational principles for solving the ground-state energy~\cite{Peruzzo2014, Wecker2015}, real time simulation~\cite{Li2017, Lau2021} and imaginary time simulation~\cite{McArdle2019, Motta2020}. A variational quantum algorithm largely depends on the ansatz, i.e. a parameterised quantum circuit. Some ansatz circuits suffer from the ``barren plateaus'' problem, which is a vanishing gradient in the parameter landscape, making the algorithm inefficient~\cite{McClean2018}. So far, a general way to construct a proper ansatz is still lacking. Applied to Hamiltonians with tens to hundreds of qubits, the performance of variational quantum algorithms on a noisy quantum computer remains an open question~\cite{Cao2019, McArdle2020}.

In this paper, we propose a hybrid non-variational quantum simulation algorithm, i.e.~the QCMC algorithm. Contrary to the QMC methods, there is no sign problem in simulating the time evolution using quantum computing. If we can delegate the calculation of the most oscillating part to quantum computing, the remaining calculations in QMC might have a very mild sign problem, or even be free from it when the entire calculation is delegated to quantum computing. To explore this possibility, we carry out the QCMC simulation by sampling random quantum circuits. Several aspects of this hybrid scheme are discussed, including implementation of the time evolution operators, the total computational complexity, the optimal sampling distribution in Monte Carlo, and the error-mitigation techniques. We show that our algorithm is polynomial on a fault-tolerant quantum computer and can reduce the variance of the Monte Carlo estimator even on a noisy quantum computer. As a subroutine of QMC, the circuit depth in quantum computing can be drastically reduced compared with the conventional Lie-Trotter-Suzuki decomposition. Therefore, our algorithm is a suitable candidate for the near-term application of quantum computing.

In the QCMC algorithm, we simulate many-body dynamics by expressing the time evolution operator in a summation form. Each term in the summation corresponds to a quantum circuit configuration. The summation formula is chosen to minimise the circuit depth and variance of the Monte Carlo estimator. We introduce two series of summation formulas based on Lie-Trotter-Suzuki product formulas~\cite{Suzuki1990, Yoshida1990}: Pauli-operator-expansion (POE) formulas and leading-order-rotation (LOR) formulas. Compared with product formulas, in our formulas the algorithmic error converges faster with the time step size $\Delta t$, at the cost of a moderately increased gate number per time step. For example, the second-order LOR formula converges as $O(\Delta t^6)$, which is even faster than the fourth-order product formula. This algorithmic error in QCMC is only due to the variance of the Monte Carlo estimator and can be reduced by increasing the sample number.

We mitigate errors in QCMC in three ways. First, our summation formulas are exact formulas of the time evolution operator for any finite time step size. The  product formulas have the decomposition error depending on $\Delta t$, which must be sufficiently small to reduce the error. Exact summation formulas allow us to take a large $\Delta t$ (i.e.~a small number of time steps) and use shallow circuits to implement QCMC. We remark that the gate number per time step is only moderately increased to implement the proposed summation formulas. Second, we use quantum error mitigation techniques to eliminate the impact of machine errors caused by noise in the quantum computer~\cite{Li2017, Temme2017, McClean2017}. We present two types of circuits: forward-backward circuits have larger depths than compact circuits but provide inherent error mitigation. Alternatively, probabilistic error cancellation is a universal way to mitigate machine errors, which enlarges the estimator variance by a factor depending on the circuit depth~\cite{Temme2017, Endo2018}. Considering probabilistic error cancellation applied to compact circuits, we can estimate the overall variance of QCMC due to both QMC and error mitigation. Third, we minimise the variance, i.e.~the statistical error, by taking the optimal time step size. We obtain the minimised variance of QCMC in the form of approximately $e^{4\gamma h_{\rm tot}t}$, where $\gamma$ is the increasing rate of the variance, $h_{\rm tot}$ characterises the magnitude of the Hamiltonian, and $t$ is the evolution time.

QCMC has a variance that depends on the rate of machine errors and achieves a quantum advantage even when the error rate is finite. For the second-order LOR formula, rate of increase of variance has the upper bound $\gamma\simeq 2.45 \epsilon^{0.82}$, where $\epsilon$ is the total gate error rate of one elementary Lie-Trotter-Suzuki product (i.e.~the first-order product for one time step). QCMC is polynomial on a fault-tolerant quantum computer because we can suppress $\epsilon$ to any small value at a polynomial cost in quantum error correction. Suppose that the variance in classical algorithms is in the same exponential form with a finite increasing rate $\gamma_{\rm c}$~\cite{Troyer2005}: the quantum algorithm surpasses the classical algorithms given an error rate of $\epsilon \lesssim (\gamma_{\rm c}/2.45)^{1/0.82}$. As an example, the rate of increase of variance in Green's function Monte Carlo taking the computational basis is $\gamma_{\rm c} = 1$ for a large class of qubit Hamiltonians. Compared with this classical algorithm, QCMC reduces the variance by several orders of magnitude even on a quantum computer with significant noise, e.g.~by a factor of approximately $4\times 10^4$ when $h_{\rm tot}t = 4$ and $\epsilon = 0.1$. As a result, the sample number required in Monte Carlo is reduced by the same factor.

In this paper, we focus on the non-variational simulation of real time evolution. With the real time simulation, we can construct quantum phase estimation circuits~\cite{OBrien2019} and eigenenergy filtering operators~\cite{Lu2021} to solve eigenstate and finite-temperature problems. The QCMC algorithm also provides a flexible tool for variational quantum algorithms. Here, we present two such examples. First, the ground state and other eigenstates are stationary and do not evolve with time, which leads to a way of ruling out fallacious solutions from the variational quantum eigensolver: if we find that the state evolves in the real time simulation, the initial state must not be an eigenstate. Second, the optimiser in the variational algorithm may get stuck in a local minimum; then, real time evolution can be used to bring the state out of the local minimum without changing the average energy. Note that by using shallow circuits in QCMC, the overall circuit combining the variational ansatz and the time evolution are still within the regime of near-term application.

This paper is organised as follows. In Sec.~\ref{sec:QMC}, we briefly review Green's function Monte Carlo and auxiliary-field Monte Carlo. In Sec.~\ref{sec:QCMC}, we sketch the QCMC algorithm. Two series of summation formulas are introduced in Sec.~\ref{sec:formulas}. Details of the QCMC algorithm are presented in the form of pseudocode in Sec.~\ref{sec:algorithm}. In Sec.~\ref{sec:circuits}, we give two types of quantum circuits (i.e.~compact circuits and forward-backward circuits) for evaluating transition amplitudes. In Sec.~\ref{sec:distribution}, we discuss the optimal distribution for generating samples in Monte Carlo. Two quantum error mitigation protocols using probabilistic error cancellation and forward-backward circuits, respectively, are discussed in Sec.~\ref{sec:qem}. The QCMC algorithm and the classical QMC algorithm are compared in Sec.~\ref{sec:QandC}. In Sec.~\ref{sec:conclusions}, we summarise the conclusions.

\section{Quantum Monte Carlo}
\label{sec:QMC}

Many applications of QMC can be formalised as computing the transition amplitude $\bra{\psi_{\rm f}} e^{iHt^*} O e^{-iHt} \ket{\psi_{\rm i}}$ given the initial state $\ket{\psi_{\rm i}}$, the final state $\ket{\psi_{\rm f}}$, and the operator $O$. Here, $H$ is the Hamiltonian, and $t$ is a real or imaginary evolution time. For example, the ground state energy of an interacting Hamiltonian can be expressed as
\begin{eqnarray}
E_{g.s.} = \lim_{t\rightarrow \infty}\frac{\langle \psi_0 | e^{-Ht/2} H e^{-Ht/2} | \psi_0 \rangle}
{\langle \psi_0 | e^{-Ht} | \psi_0 \rangle},
\end{eqnarray}
where $|\psi_0\rangle$ is a trial ground state, which has a large overlap with the true ground state.

A canonical approach is Green's function Monte Carlo~\cite{Carlson2015}, in which the transition amplitude is expressed in the path-integral form:
\begin{eqnarray}
&& \bra{\psi_{\rm f}} e^{iHt^*} O e^{-iHt} \ket{\psi_{\rm i}} \notag \\
&=& \int_{\bfr_0,\ldots,\bfr_N,\bfr_0',\ldots,\bfr_N'} d\bfr_0\cdots d\bfr_N d\bfr_0'\cdots d\bfr_N' \notag \\
&&\braket{\psi_{\rm f}}{\bfr_0'} \bra{\bfr_0'}e^{iH\frac{t^*}{N}}\ket{\bfr_1'} \cdots \bra{\bfr_{N-1}'}e^{iH\frac{t^*}{N}}\ket{\bfr_N'} \bra{\bfr_N'}O\ket{\bfr_N} \notag \\
&&\times \bra{\bfr_N}e^{-iH\frac{t}{N}}\ket{\bfr_{N-1}} \cdots \bra{\bfr_1}e^{-iH\frac{t}{N}}\ket{\bfr_0} \braket{\bfr_0}{\psi_{\rm i}},
\label{eq:GFQMC}
\end{eqnarray}
where $\{\ket{\bfr}\}$ is an orthonormal basis of the Hilbert space and $N$ is the number of time steps. The path integral is performed numerically using Monte Carlo methods.

Auxiliary-field Monte Carlo is another important approach of QMC~\cite{Blankenbecler1981, Lee2009}, which is characterized by the decomposition of particle-particle interactions into interactions of particles with a group of auxiliary fields, i.e.
\begin{eqnarray}
e^{-iH\Delta t} \simeq \int ds A(s,\Delta t).
\end{eqnarray}
Here, $A(s,\Delta t)$ is an operator depending on the auxiliary field $s$. Then, the transition amplitude is expressed as
\begin{eqnarray}
&& \bra{\psi_{\rm f}} e^{iHt^*} O e^{-iHt} \ket{\psi_{\rm i}} \notag \\
&=& \int ds_1\cdots ds_N ds_1'\cdots ds_N' \bra{\psi_{\rm f}} A(s_1',-\Delta t^*) \cdots  \notag \\
&&\times A(s_N',-\Delta t^*) O A(s_N,\Delta t) \cdots A(s_1,\Delta t) \ket{\psi_{\rm i}}.
\label{eq:AFQMC}
\end{eqnarray}
The operator $A(s,\Delta t)$ is chosen such that $\bra{\psi_{\rm f}}\cdots \ket{\psi_{\rm i}}$ in the integral can be evaluated on a classical computer.

In diagrammatic QMC, the time evolution amplitudes are expressed as perturbative expansions~\cite{Houcke2010, Houcke2012, Cohen2015, Bertrand2019}. Suppose that the contribution of an $m$-th-order term is $D(\xi_m,x_1,\ldots,x_m)$: the transition amplitude is a summation of integrals in the form
\begin{eqnarray}
&& \bra{\psi_{\rm f}} e^{iHt^*} O e^{-iHt} \ket{\psi_{\rm i}} \notag \\
&=& \sum_{m=0}^{\infty} \sum_{\xi_m} \int dx_1\cdots dx_m D(\xi_m,x_1,\ldots,x_m),
\end{eqnarray}
where $\xi_m$ is the index of the term and the $x$ are the temporal and spatial coordinates to be integrated. These terms can be represented by Feynman diagrams. In these models, we can develop similar quantum algorithms, in which both the non-interacting time evolution and the interaction vertices can be implemented as a series of operators that can be evaluated on a quantum computer.

It often occurs that the amplitude $q = \bra{\psi_{\rm f}}\cdots \ket{\psi_{\rm i}}$ as a function of ${\bfr}$ in Eq.~(\ref{eq:GFQMC}) or as a function of $s$ in Eq.~(\ref{eq:AFQMC}) is not positive definite. In this case, we have to use the reweighting procedure by splitting $q$ into its modulus and phase, i.e.~$q = |q|e^{i\theta_q}$, and sample according to a probability distribution $P \propto |q|$. The expectation value of the remaining phase $\langle e^{i\theta_q} \rangle$ indicates the degree of the sign problem and if it is much smaller than $1$ then the sign problem is severe. In many QMC simulations, this phase goes to zero exponentially for a large system volume or particle number, which signifies a very bad sign problem.

In some special cases, the sign problem is only induced by part of the integral variables. In other words, the amplitude $q$ is a highly oscillating function of some variables and a smooth function of the others. This usually occurs when the system is protected by an approximate symmetry. For example, for fermion systems with equal numbers of up and down spins, a spin-independent attractive interaction respecting the SU(2) spin symmetry does not induce the sign problem. In more general problems, the realistic interaction might be dominated by such a ``good'' component, while other ``bad'' components play a minor role but induce most of the sign problem. A typical example is the nuclear force, which is approximately independent of spin and isospin at low energy~\cite{Lee2020}. The spin-isospin dependent components and the Coulomb force only contribute a small portion of the total nuclear binding energy but introduce strong a sign problem in the auxiliary field Monte Carlo calculations. Usually, these interactions can be simulated using the coupling constant extrapolation method~\cite{Lahde2015}, perturbation theory~\cite{Epelbaum2014} or the eigenvector continuation method~\cite{Frame2018, Konig2020, Sarkar2021}, at the cost of additional uncertainties.

The above problem has an alternative solution in the quantum computing era. As a quantum computer can calculate the amplitude $q$ with the same complexity regardless of the form of the interaction, we can use the quantum computer to simulate interactions causing the sign problem, while leaving the smooth high-dimensional integrals to the classical Monte Carlo solver. For example, in the auxiliary-field Monte Carlo simulation of atomic nuclei~\cite{Lee2009}, we can simulate the repulsive Coulomb force using quantum computing. In this paper, we introduce such a hybrid simulation scheme and establish a general framework for future work in this direction.

\section{Quantum-circuit Monte Carlo}
\label{sec:QCMC}

\begin{figure}[tbp]
\begin{center}
\includegraphics[width=1\linewidth]{\figpath/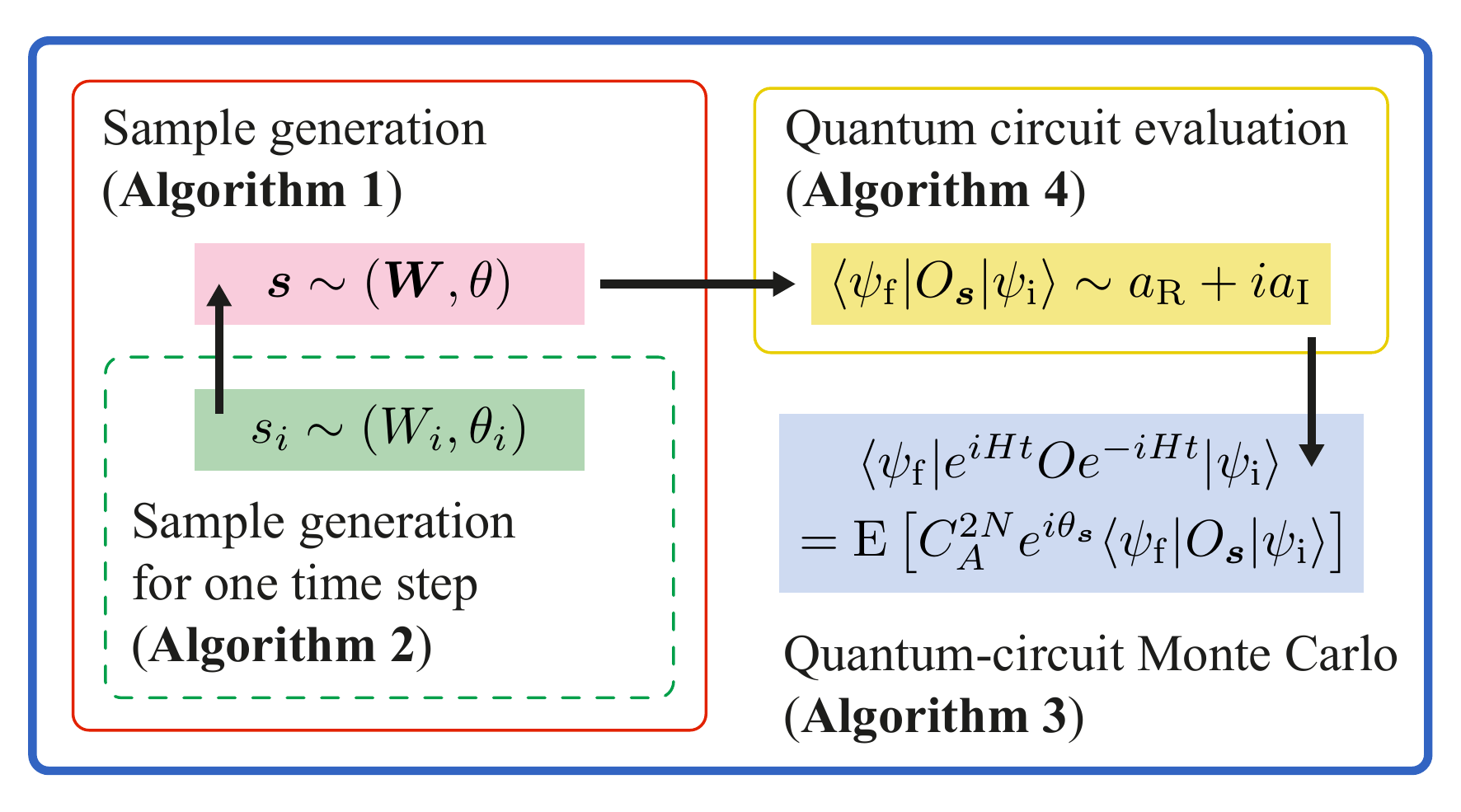}
\caption{
A schematic diagram of the quantum-circuit Monte Carlo algorithm. The quantum computer evaluates $\bra{\psi_{\rm f}} O_{\bfs} \ket{\psi_{\rm i}}$ according to samples $\bfs$ generated by the classical computer. The final estimate of the transition amplitude $\bra{\psi_{\rm f}} e^{iHt} O e^{-iHt} \ket{\psi_{\rm i}}$ is the empirical mean of results from the quantum computer up to a factor.
}
\label{fig:flowchart}
\end{center}
\end{figure}

To implement QMC using a quantum computer, we replace the integral over the auxiliary field with a summation over unitary operators. The time evolution operator is expressed in the summation form
\begin{eqnarray}
e^{-iH\Delta t} = \sum_{s} c(s) U(s),
\end{eqnarray}
where the $U(s)$ are unitary operators and the $c(s)$ are complex coefficients. For real time evolution, approximate summation formulas have been proposed, including truncated Taylor expansion~\cite{Berry2015, Meister2020} and linear combinations of Lie-Trotter-Suzuki products~\cite{Childs2012, Faehrmann2021}. In this paper, we propose exact summation formulas of the real time evolution operator (see Sec.~\ref{sec:formulas}). Note that we can also construct the imaginary time evolution operator as a summation of unitary operators and construct any operator in the limit that the $U(s)$ form a complete basis of the operator space. By combining quantum circuits and the Monte Carlo method, our exact formulas can be implemented for any finite time step size $\Delta t$. In quantum circuits, the gate number per time step is only moderately increased upon the Lie-Trotter-Suzuki product (see Sec.~\ref{sec:circuits}) and we can minimise the number of time steps $N$ by maximising $\Delta t$. Because of the minimised circuit depth, which is proportional to $N$, our formulas are practical on noisy quantum computers without fault tolerance.

With the summation expression of the time evolution operator, the transition amplitude in the path-integral form becomes
\begin{eqnarray}
&& \bra{\psi_{\rm f}} e^{iHt} O e^{-iHt} \ket{\psi_{\rm i}} \notag \\
&=& \sum_{s_1,\ldots,s_N,s_1',\ldots,s_N'} \left( \prod_{i=1}^N c(s_i)c(s_i')^* \right) \bra{\psi_{\rm f}} O_{\bfs} \ket{\psi_{\rm i}}
\label{eq:QCMC}
\end{eqnarray}
where $\bfs = (s_1,\ldots,s_N,s_1',\ldots,s_N')$ and
\begin{eqnarray}
O_{\bfs} = U(s_1')^\dag \cdots U(s_N')^\dag O U(s_N) \cdots U(s_1).
\end{eqnarray}
One can realise a summation formula either by using a deterministic circuit~\cite{Berry2015, Meister2020, Childs2012} or sampling random circuits~\cite{Faehrmann2021, Campbell2019}. To minimise the circuit depth, we compute the transition amplitude using random circuits: we sample random unitary operators (i.e.~the parameter $\bfs$) on the classical computer, evaluate $\bra{\psi_{\rm f}} O_{\bfs} \ket{\psi_{\rm i}}$ on the quantum computer and then compute the path-integral summation using the Monte Carlo method on the classical computer. See Fig.~\ref{fig:flowchart} for a schematic diagram of the QCMC algorithm and see Sec.~\ref{sec:algorithm} for details.

Without fault tolerance, we use error mitigation techniques to eliminate errors in quantum circuits. In the quantum error mitigation based on quasi-probability decomposition (i.e.~probabilistic error cancellation)~\cite{Temme2017, Endo2018}, each unitary circuit for evaluating $\bra{\psi_{\rm f}} O_{\bfs} \ket{\psi_{\rm i}}$ is decomposed into a linear combination of noisy circuits. Then, the overall algorithm includes Monte Carlo summations over unitary operators and also noisy circuits. Details of the error mitigation are given in Sec.~\ref{sec:qem}. Using our exact formulas of the time evolution operator and assuming that quasi-probability decompositions are also exact, the sampling noise in Monte Carlo is the only source of error in our algorithm.

\subsection*{Sampling noise and normalisation factor}

The Monte Carlo summation has a finite variance depending on the sampling approach. To compute the transition amplitude in Eq.~(\ref{eq:QCMC}), we randomly generate samples of $\bfs$ with a probability distribution $P(\bfs)$. According to the importance sampling, the variance is minimised by taking the optimal distribution
\begin{eqnarray}
P(\bfs) \propto \absLR{\left( \prod_{i=1}^N c(s_i)c(s_i')^* \right) \bra{\psi_{\rm f}} O_{\bfs} \ket{\psi_{\rm i}}}.
\label{eq:optimalPRO}
\end{eqnarray}
Implementation of the optimal distribution requires knowledge of $\absLR{\bra{\psi_{\rm f}} O_{\bfs} \ket{\psi_{\rm i}}}$.

In this paper, we focus on a practical suboptimal distribution
\begin{eqnarray}
P(\bfs) = \absLR{\prod_{i=1}^N c(s_i)c(s_i')^*} / C_A^{2N},
\label{eq:pro}
\end{eqnarray}
where the normalisation factor $C_A = \sum_s \abs{c(s)}$ determines the variance. Taking the suboptimal distribution, the transition amplitude $\bra{\psi_{\rm f}} e^{iHt} O e^{-iHt} \ket{\psi_{\rm i}}$ is the expected value of $C_A^{2N} e^{i\theta_{\bfs}}\bra{\psi_{\rm f}} O_{\bfs} \ket{\psi_{\rm i}}$, where
\begin{eqnarray}
\theta_{\bfs} = \arg\left(\prod_{i=1}^N c(s_i)c(s_i')^*\right).
\end{eqnarray}
Formally, we have
\begin{eqnarray}
&& \bra{\psi_{\rm f}} e^{iHt} O e^{-iHt} \ket{\psi_{\rm i}} = \mathrm{E}\left[ C_A^{2N} e^{i\theta_{\bfs}} \bra{\psi_{\rm f}} O_{\bfs} \ket{\psi_{\rm i}} \right] \notag \\
&=& \sum_{\bfs} P(\bfs) C_A^{2N} e^{i\theta_{\bfs}} \bra{\psi_{\rm f}} O_{\bfs} \ket{\psi_{\rm i}}.
\end{eqnarray}

Taking the suboptimal distribution, the estimator of $\bra{\psi_{\rm f}} e^{iHt} O e^{-iHt} \ket{\psi_{\rm i}}$ is
\begin{eqnarray}
\hat{A} = C_A^{2N} \meanLR{e^{i\theta_{\bfs}}\bra{\psi_{\rm f}} O_{\bfs} \ket{\psi_{\rm i}}}_{N_{\rm s}}.
\end{eqnarray}
Here, $\meanLR{\bullet}_{N_{\rm s}}$ denotes the empirical mean taken over $N_{\rm s}$ samples of $\bfs$. The variance of the estimator is
\begin{eqnarray}
\mathrm{Var}\left(\hat{A}\right) = \frac{1}{N_{\rm s}} C_A^{4N} \mathrm{Var}\left(e^{i\theta_{\bfs}}\bra{\psi_{\rm f}} O_{\bfs} \ket{\psi_{\rm i}}\right).
\end{eqnarray}
When $O$ is a unitary operator, $\abs{\bra{\psi_{\rm f}} O_{\bfs} \ket{\psi_{\rm i}}} \leq 1$, and the variance has the upper bound
\begin{eqnarray}
\mathrm{Var}\left(\hat{A}\right) \leq \frac{1}{N_{\rm s}} C_A^{4N}.
\end{eqnarray}

In our QCMC algorithm, we use the circuits given in Sec.~\ref{sec:circuits} to evaluate $\bra{\psi_{\rm f}} O_{\bfs} \ket{\psi_{\rm i}}$. Each quantum circuit reports a probabilistic binary outcome, the expected value of which is either the real or imaginary part of $e^{i\theta_{\bfs}}\bra{\psi_{\rm f}} O_{\bfs} \ket{\psi_{\rm i}}$. We find that the suboptimal distribution (which is suboptimal when we can deterministically evaluate $\bra{\psi_{\rm f}} O_{\bfs} \ket{\psi_{\rm i}}$) is actually the optimal distribution for the probabilistic evaluation without prior knowledge of $\absLR{\bra{\psi_{\rm f}} O_{\bfs} \ket{\psi_{\rm i}}}$ (see~\ref{sec:distribution}). Accordingly, the minimum variance is
\begin{eqnarray}
\mathrm{Var}\left(\hat{A}\right) = \frac{1}{M_{\rm tot}} \left(2C_A^{4N} - \absLR{\bra{\psi_{\rm f}} e^{iHt} O e^{-iHt} \ket{\psi_{\rm i}}}^2\right),~~~
\label{eq:var}
\end{eqnarray}
where $2M_{\rm tot}$ is the total number of quantum circuit shots, and each shot is an implementation of the circuit that returns one binary measurement outcome.

We find that ideally $C_A = 1$, i.e.~the variance does not increase with the number of time steps. This limit can be approached on a fault-tolerant quantum computer: we take a sufficiently small $\Delta t$, $c(1) \simeq 1$, $U(1) \simeq e^{-iH\Delta t}$ is a Lie-Trotter-Suzuki product, and terms with $s>1$ are negligible. On a noisy quantum computer, $C_A$ is always greater than one. A large part of our effort is devoted to minimising $C_A$, in order to reduce the variance.

\section{Summation formulas of time evolution operators}
\label{sec:formulas}

\begin{table*}[tbp]
\begin{center}
\begin{tabular}{|c|c|c|c|c|}
\hline
Pauli-operator-expansion formulas & \multicolumn{4}{c|}{$C_A = 1+C_L+C_T$} \\
\hline
Leading-order-rotation formulas & \multicolumn{4}{c|}{$C_A = \sqrt{1+C_L^2}+C_T$} \\
\hline
High-order contribution $C_T$ & \multicolumn{4}{c|}{$e^{\lambda x} - \sum_{k=0}^{2l+1}\frac{1}{k!}(\lambda x)^k$} \\
\hline
\hline
$l$ (order of formula) & $0$ & $1$ & $2$ & $2m$ \\
\hline
$\lambda$ & $1$ & $2$ & $2$ & $1+\prod_{k=2}^m \left(4rp_{r,k} - 1\right)$ \\
\hline
Leading-order contribution $C_L$ & ~$x$~ & ~$\frac{1}{2}(2x)^2 + \frac{1}{6}(2x)^3$~ & ~$\frac{1}{6}(2x)^3 + \frac{1}{120}(2x)^5$~ & ~$\sum_{k=m}^{2m} \frac{1}{(2k+1)!}(\lambda x)^{2k+1}$~ \\
\hline
Simplified leading-order contribution $C_L$ & & $< \frac{1}{2}x^2 + \frac{1}{6}(2x)^3$ & $< \frac{1}{18}x^3 + \frac{1}{120}(2x)^5$ & \\
\hline
\end{tabular}
\end{center}
\caption{
Normalisation factors of summation formulas. In the table, $x \equiv h_{\rm tot}\Delta t$.
}
\label{table:CA}
\end{table*}

We look for summation formulas satisfying the following criteria:
\begin{itemize}
\item The unitary operators $U(s)$ are easy to implement using elementary quantum gates, in order to reduce the gate number.
\item The normalisation factor $C_A$ is minimised.
\item Samples of $\bfs$ can be efficiently generated on a classical computer according to the distribution in Eq.~(\ref{eq:pro}).
\end{itemize}

We propose two types of summation formulas in this paper as examples of the general approach. By adding Pauli operators to Lie-Trotter-Suzuki products, we obtain POE formulas. For an $l$th-order product formula, the corresponding POE summation formula has the normalisation factor $C_A = 1+O(\Delta t^{l+1})$. By replacing leading-order Pauli operators with rotation operators, we obtain LOR formulas and the normalisation factor is reduced to $C_A = 1+O(\Delta t^{2l+2})$.

In the following, we first discuss Lie-Trotter-Suzuki product formulas and then introduce our summation formulas.

\subsection{Product formulas}

In this section, we review Lie-Trotter-Suzuki product formulas~\cite{Suzuki1990, Yoshida1990} and discuss some properties that are important for our discussion. Given the Hamiltonian $H = \sum_{j=1}^M H_j$, where the $H_j$ are Hermitian operators, the first-order formula reads
\begin{eqnarray}
S_{1}(\Delta t) = e^{-iH_M\Delta t} \cdots e^{-iH_1\Delta t} = e^{-iH\Delta t} + O(\Delta t^2).~~~
\label{eq:S1}
\end{eqnarray}
Higher-order formulas are defined recursively for any positive integer $m$ by
\begin{eqnarray}
S_{2m}(\Delta t) &=& K_{2m}(-\Delta t)^\dag K_{2m}(\Delta t) \notag \\
&=& e^{-iH\Delta t}+ O(\Delta t^{2m+1}),
\end{eqnarray}
where $K_{2}(\Delta t) = S_{1}(\frac{\Delta t}{2})$,
\begin{eqnarray}
K_{2m}(\Delta t) = K_{2m-2}\left((1-2rp_{r,m})\Delta t\right) S_{2m-2}\left(p_{r,m}\Delta t\right)^r~~~~~
\end{eqnarray}
when $m>1$, and $p_{r,m} = \left[2r - (2r)^{\frac{1}{2m+1}}\right]^{-1}$. Here, $r$ can be any positive integer. $S_{2m}$ is a product of $2r+1$ $S_{2m-2}$ operators.

For the first-order formula, we define the correction operator
\begin{eqnarray}
V_{1}(\Delta t) \equiv e^{-iH\Delta t} S_{1}(\Delta t)^\dag = e^{-i\sum_{k=2}^{\infty} R_1^{(k)}\Delta t^k},
\label{eq:CO1}
\end{eqnarray}
where $R_1^{(k)}$ are operators that are independent of $\Delta t$. Because $V_{1}(\Delta t)$ is unitary for all real $\Delta t$, all $R_1^{(k)}$ are Hermitian operators. Then,
\begin{eqnarray}
V_{1}(\Delta t) = \openone - iL_{1}(\Delta t) + O(\Delta t^4),
\end{eqnarray}
where the leading-order operator
\begin{eqnarray}
L_{1}(\Delta t) = R_1^{(2)}\Delta t^2 + R_1^{(3)}\Delta t^3
\end{eqnarray}
is Hermitian. Later, we show that the Hermitian leading-order operator is important for minimising the normalisation factor $C_A$.

For higher-order formulas, the correction operators are
\begin{eqnarray}
V_{2m}(\Delta t) &\equiv & K_{2m}(-\Delta t) e^{-iH\Delta t} K_{2m}(\Delta t)^\dag \notag \\
&=& e^{-i\sum_{k=2m+1}^{\infty} R_{2m}^{(k)}\Delta t^k},
\label{eq:CO2m}
\end{eqnarray}
where $R_{2m}^{(k)}$ are Hermitian operators that are independent of $\Delta t$. Because of the symmetric form, $V_{2m}(\Delta t) = V_{2m}(-\Delta t)^\dag$ for all real $\Delta t$, and $R_{2m}^{(k)} = 0$ for all even $k$~\cite{Yoshida1990}. Then,
\begin{eqnarray}
V_{2m}(\Delta t) = \openone - iL_{2m}(\Delta t) + O(\Delta t^{4m+2}),
\end{eqnarray}
where the leading-order operator
\begin{eqnarray}
L_{2m}(\Delta t) = \sum_{k=m}^{2m} R_{2m}^{(2k+1)}\Delta t^{2k+1}
\end{eqnarray}
is Hermitian. For the second-order formula,
\begin{eqnarray}
L_{2}(\Delta t) = R_{2}^{(3)}\Delta t^3 + R_{2}^{(5)}\Delta t^5.
\end{eqnarray}

\subsection{Summation formulas}

To simplify the quantum circuits, we work with Pauli operators $\bfP_n = \{I,X,Y,Z\}^{\otimes n}$ as the basis of matrix space, where $n$ is the number of qubits. Without loss of generality, we assume that each term of the Hamiltonian is a Pauli operator, i.e.~$H_j = h_j \sigma_j$, where $\sigma_j \in \bfP_n$, and $h_j$ is a real coefficient. We define $h_{\rm tot} \equiv \sum_j \abs{h_j}$, which characterises the magnitude of the Hamiltonian.

Given the time evolution operator, there exist many different summation formulas $e^{-iH\Delta t} = \sum_s c(s) U(s)$. Each formula represents a sampling protocol in Monte Carlo. For example,
\begin{eqnarray}
e^{-iH\Delta t} = \sum_{\sigma \in \bfP_n} 2^{-n} \Tr\left(\sigma e^{-iH\Delta t}\right) \sigma.
\end{eqnarray}
Such a formula is impractical, because the computing of the coefficients $\Tr\left(\sigma e^{-iH\Delta t}\right)$ on a classical computer is usually difficult when $n$ is large.

For the practical implementation, we express the time evolution operator in the form
\begin{eqnarray}
e^{-iH\Delta t} = K_L V K_R,
\end{eqnarray}
where $K_L$ and $K_R$ are unitary operators in the Lie-Trotter-Suzuki product form, and $V$ is the correction operator, see Eqs.~(\ref{eq:CO1}) and (\ref{eq:CO2m}). We apply the Taylor expansion to the correction operator to obtain the summation formula. We divide the Taylor expansion into three parts, $V = \openone -iL + T$, where $L$ is the leading-order operator, and $T$ is the high-order operator. The normalisation factor of a POE summation formula is $C_A = 1 + C_L + C_T$, where $C_L$ and $C_T$ are contributions of $L$ and $T$, respectively. The normalisation factor of a LOR summation formula is $C_A = \sqrt{1 + C_L^2} + C_T$. The normalisation factors of all the formulas are summarised in Table~\ref{table:CA}.

\subsubsection{Zeroth-order Pauli-operator-expansion formula}

The direct Taylor expansion of the time evolution operator gives the zeroth-order summation formula
\begin{eqnarray}
V_{0}(\Delta t) = e^{-iH\Delta t} = \openone - iL_{0}(\Delta t) + T_{0}(\Delta t),
\end{eqnarray}
where the Hermitian leading-order operator is
\begin{eqnarray}
L_{0}(\Delta t) = \sum_{j} h_j\Delta t\sigma_j,
\end{eqnarray}
and the high-order operator is
\begin{eqnarray}
T_{0}(\Delta t) = \sum_{k=2}^{\infty} \sum_{j_1,\ldots ,j_k = 1}^M \frac{\prod_{a=1}^k \left(-ih_{j_a}\Delta t\right)}{k!} \sigma_{j_k}\cdots \sigma_{j_1}.~~~
\end{eqnarray}
The normalisation factor is given by $C_L = h_{\rm tot}\Delta t$ and $C_T = e^{h_{\rm tot}\Delta t} - (1 + h_{\rm tot}\Delta t)$.

\subsubsection{First-order Pauli-operator-expansion formula}

According to the first-order product formula, we express the time evolution operator as
\begin{eqnarray}
e^{-iH\Delta t} = V_{1}(\Delta t) S_{1}(\Delta t).
\end{eqnarray}
We obtain the summation formula by applying the Taylor expansion to each exponential in the correction operator,
\begin{eqnarray}
V_{1}(\Delta t) &=& e^{-iH\Delta t} e^{ih_1\sigma_1\Delta t} \cdots e^{ih_M\sigma_M\Delta t} \notag \\
&=& \openone - iL_{1}(\Delta t) + T_{1}(\Delta t),
\end{eqnarray}
where
\begin{eqnarray}
L_{1}(\Delta t) &=& iF_{1}^{(2)}(\Delta t) + iF_{1}^{(3)}(\Delta t), \\
T_{1}(\Delta t) &=& \sum_{k=4}^{\infty} F_{1}^{(k)}(\Delta t),
\end{eqnarray}
and
\begin{eqnarray}
F_{1}^{(k')}(\Delta t) &=& \sum_{k,k_1,\ldots, k_M=0}^{\infty} \sum_{j_1,\ldots ,j_k = 1}^M \delta_{k',k+\sum_{j=1}^M k_j} \notag \\
&&\times \frac{\prod_{a=1}^k \left(-ih_{j_a}\Delta t\right)}{k!} \left[\prod_{j=1}^M \frac{(ih_{j}\Delta t)^{k_j}}{k_j!} \right] \notag \\
&&\times \sigma_{j_k}\cdots \sigma_{j_1} \sigma_1^{k_1}\cdots \sigma_M^{k_M}.
\end{eqnarray}
Note that the first term in $L_{1}$ is $O(\Delta t^2)$ according to discussions on product formulas. For the first-order formula, the normalisation factor is given by $C_L = \frac{1}{2}\left(2h_{\rm tot}\Delta t\right)^2 + \frac{1}{6}\left(2h_{\rm tot}\Delta t\right)^3$ and $C_T = e^{2h_{\rm tot}\Delta t} - \sum_{k=0}^3\frac{1}{k!}\left(2h_{\rm tot}\Delta t\right)^k$.

\subsubsection{Second-order Pauli-operator-expansion formula}

Similar to the first-order formula, according to the second-order product formula, we express the time evolution operator as
\begin{eqnarray}
e^{-iH\Delta t} = S_{1}(-\frac{\Delta t}{2})^\dag V_{2}(\Delta t) S_{1}(\frac{\Delta t}{2}).
\end{eqnarray}
The Taylor expansion of the correction operator reads
\begin{eqnarray}
V_{2}(\Delta t) &=& \openone - iL_{2}(\Delta t) + T_{2}(\Delta t),
\end{eqnarray}
where
\begin{eqnarray}
L_{2}(\Delta t) &=& iF_{2}^{(3)}(\Delta t) + iF_{2}^{(5)}(\Delta t), \label{eq:L2} \\
T_{2}(\Delta t) &=& \sum_{k=6}^{\infty} F_{2}^{(k)}(\Delta t),
\end{eqnarray}
and
\begin{eqnarray}
F_{2}^{(k')}(\Delta t) &=& \sum_{k,k_1,\ldots, k_1',\ldots =0}^{\infty} \sum_{j_1,\ldots ,j_k = 1}^M \delta_{k',k+\sum_{j=1}^M (k_j+k_j')} \notag \\
&&\times  \frac{\prod_{a=1}^k \left(-ih_{j_a}\Delta t\right)}{k!} \left[\prod_{j=1}^M \frac{(ih_{j}\Delta t/2)^{k_j+k_j'} }{k_j!k_j'!}\right] \notag \\
&&\times \sigma_M^{k_M'}\cdots \sigma_1^{k_1'} \sigma_{j_k}\cdots \sigma_{j_1} \sigma_1^{k_1}\cdots \sigma_M^{k_M}.
\end{eqnarray}
According to discussions on product formulas, $L_{2}$ only contain $\Delta t^3$ and $\Delta t^5$ terms. For the second-order formula, the normalisation factor is given by $C_L = \frac{1}{6}\left(2h_{\rm tot}\Delta t\right)^3 + \frac{1}{120}\left(2h_{\rm tot}\Delta t\right)^5$ and $C_T = e^{2h_{\rm tot}\Delta t} - \sum_{k=0}^5\frac{1}{k!}\left(2h_{\rm tot}\Delta t\right)^k$.

\subsubsection{Higher-order Pauli-operator-expansion formulas}

For the $2m$th-order formula, we express the time evolution operator as
\begin{eqnarray}
e^{-iH\Delta t} = K_{2m}(-\Delta t)^\dag V_{2m}(\Delta t) K_{2m}(\Delta t).
\end{eqnarray}
Then, we can obtain the POE summation formula by applying a Taylor expansion to each exponential in the correction operator $V_{2m}$, similar to the first- and second-order formulas. The normalisation factor of the $2m$th-order formula is given by $C_L = \sum_{k=m}^{2m} \frac{1}{(2k+1)!}\left(\lambda h_{\rm tot}\Delta t\right)^{2k+1}$ and $C_T = e^{\lambda h_{\rm tot}\Delta t} - \sum_{k=0}^{4m+1}\frac{1}{k!}\left(\lambda h_{\rm tot}\Delta t\right)^k$. Here, the factor $\lambda = 1+\prod_{k=2}^m \left(4rp_{r,k} - 1\right)$ is due to the backward evolution with the time $(1-2rp_{r,m})\Delta t$ in the product formula.

\subsubsection{Simplified leading-order operators}

By combining terms with the same Pauli operator in the summation formula, we can reduce the normalisation factor. For example, if both $\alpha \sigma$ and $-\alpha \sigma$ exist in the summation formula, the contribution to the normalisation factor is $2\abs{\alpha}$, which is reduced to zero after combining like terms. We apply this approach to $F_{1}^{(2)}$ and $F_{2}^{(3)}$ in $L_{1}$ and $L_{2}$, respectively, to minimise the dominant contribution to the normalisation factor. See Appendix~\ref{app:LOT} for the simplified expressions of $F_{1}^{(2)}$ and $F_{2}^{(3)}$. As a result, the leading-order contributions are reduced to $C_L < \frac{1}{2}\left(h_{\rm tot}\Delta t\right)^2 + \frac{1}{6}\left(2h_{\rm tot}\Delta t\right)^3$ in the first-order formula and $C_L < \frac{1}{18}\left(h_{\rm tot}\Delta t\right)^2 + \frac{1}{120}\left(2h_{\rm tot}\Delta t\right)^5$ in the second-order formula.

\subsubsection{Leading-order-rotation formulas}

The leading-order operator $L_{l}$ is Hermitian, which allows us to reduce its contribution to the normalisation factor $C_A$ from $O(\Delta t^{l+1})$ to $O(\Delta t^{2l+2})$. We suppose that the Pauli-operator summation form of $L_{l}$ is
\begin{eqnarray}
L_{l} = \sum_u \alpha_u \tau_u,
\end{eqnarray}
where the $\tau_u\in \bfP_n$ are Pauli operators. Here, all $\alpha_u$ are real because $L_{l}$ is Hermitian, which is the key to LOR formulas. To minimise the normalisation factor, we express the leading-order terms as a summation of rotation operators,
\begin{eqnarray}
\openone - iL_{l} = \sum_u \beta_u e^{-i {\rm sgn}(\alpha_u) \phi \tau_u},
\end{eqnarray}
where $\phi = \arctan(C_L)$, $\beta_u = \abs{\alpha_u}/\sin\phi$ and $C_L = \sum_u \abs{\alpha_u}$.

The normalisation factor contributed by $\openone - iL_{l}$ is $1+C_L$ in POE formulas, which is reduced to $\sum_i \abs{\beta_u} = C_L/\sin\phi = \sqrt{1+C_L^2} \simeq 1+C_L^2/2$ in LOR formulas. Note that $C_L = O(\Delta t^{l+1})$ and $C_T = O(\Delta t^{2l+2})$. By using LOR formulas, we reduce the normalisation factor $C_A$ from $1+O(\Delta t^{l+1})$ to $1+O(\Delta t^{2l+2})$.

We have introduced all of our summation formulas. We remark that our summation formulas are used for sampling random $U(s)$ rather than sampling quantum operations~\cite{Campbell2019}, which corresponds to a summation of completely positive maps instead of operators.

\subsection{Comparison between formulas}

Now, we compare different formulas of the time evolution operator in the fault-tolerance limit, i.e.~gate errors are negligible. In this case, we can use deep quantum circuits to implement the formulas and take a sufficiently small time step size $\Delta t$. We leave the discussions on noisy quantum computing to Secs.~\ref{sec:qem} and \ref{sec:QandC}.

When gate errors are negligible, sampling noise is the only source of error for our exact summation formulas. The error due to sampling noise is of approximately $\frac{1}{\sqrt{N_{\rm s}}}C_A^{2N}$. Therefore, the error for the $l$th-order POE formula is of approximately $\frac{1}{\sqrt{N_{\rm s}}} + \frac{2N}{\sqrt{N_{\rm s}}}O(\Delta t^{l+1})$ and the error for the $l$th-order LOR formula is of approximately $\frac{1}{\sqrt{N_{\rm s}}} + \frac{2N}{\sqrt{N_{\rm s}}}O(\Delta t^{2l+2})$.

For Lie-Trotter-Suzuki product formulas, there are two sources of error: the error due to finite $\Delta t$, i.e.~the formulas are approximate, and the error due to sampling noise. The error due to finite $\Delta t$ is systematic and cannot be reduced by increasing the number of samples. For the $l$-th order product formula, the error is of approximately $\frac{1}{\sqrt{N_{\rm s}}} + N O(\Delta t^{l+1})$, where the first term is due to the sampling noise and the second term is due to the finite $\Delta t$. We note that on a fault-tolerant quantum computer, we can use amplitude amplification to accelerate the evaluation of an amplitude of the wave function~\cite{Brassard2002}. Amplitude amplification can be applied to product formulas; how to apply it to our summation formulas is an open question.

We find that for the same order of formulas, our summation formulas have a smaller error than product formulas, due to the factor $\frac{1}{\sqrt{N_{\rm s}}}$ in the $\Delta t$ term and the increased exponent of $\Delta t$ (for LOR formulas). The reduced error is at the cost of an increased gate number per time step: to implement our formulas, we need to add a correction operator to the Lie-Trotter-Suzuki product for each time step. A correction operator is either a Pauli operator $\sigma$ or a rotation operator in the form $e^{-i\phi\sigma}$. In Sec.~\ref{sec:correction}, we show that implementation of the correction operator for POE and LOR formulas requires at most $n$ and $4n$ controlled-NOT gates, respectively, on an all-to-all qubit network ($4n-3$ and $8n-4$ gates, respectively, on a linear qubit network). Here, $n$ is the qubit number. Unless the Hamiltonian has the simplest structure, such as the one-dimensional quantum Ising model, it is reasonable to assume that the gate number for the first-order Lie-Trotter-Suzuki product $S_1$ is more than $2n$. Therefore, the gate number increment in each time step is moderate.

The linear combination of Lie-Trotter-Suzuki products can efficiently reduce the error due to finite $\Delta t$~\cite{Childs2012, Faehrmann2021}. The simplest example is $e^{-iH\Delta t} = \frac{4}{3} S_{2}(\Delta t/2)^2 - \frac{1}{3} S_{2}(\Delta t) + O(\Delta t^5)$. We can find that the error for our second-order LOR formula converges faster as $O(\Delta t^6)$, and the gate number is smaller compared with $S_{2}^2$ (assuming that the gate number for one $S_2$ is larger than a correction operator).

\section{Algorithm}
\label{sec:algorithm}

The algorithm consists of three phases. First, the classical computer generates samples of $\bfs$ according to the distribution given by Eq.~(\ref{eq:pro}) and composes corresponding quantum circuits. Second, the quantum computer implements circuits to evaluate $\bra{\psi_{\rm f}} O_{\bfs} \ket{\psi_{\rm i}}$. Finally, with results from the quantum computer, the classical computer calculates the expected value of $e^{i\theta_{\bfs}}\bra{\psi_{\rm f}} O_{\bfs} \ket{\psi_{\rm i}}$ and returns the final estimate of the transition amplitude $\bra{\psi_{\rm f}} e^{iHt} O e^{-iHt} \ket{\psi_{\rm i}}$.

In this section, we present the first and final phases of the algorithm, which are implemented on the classical computer. We leave details of the second phase, i.e., the quantum computing, to Sec.~\ref{sec:circuits}. We focus on second-order summation formulas and the algorithms for the other summation formulas are similar.

Our algorithm has some implicit connections to the diagrammatic Monte Carlo, in which the Feynman diagrams represent the perturbative expansions for interacting amplitudes. Similarly, the summation formulas in our algorithm are perturbativelike expansions around Lie-Trotter-Suzuki products. In our case, each term represents a path in the Hilbert space defined by the unitary operator $U(s)$ instead of $\xi_m$ and $x$ and these paths constitute the time evolution, which resembles the path-integral picture. This connection may be further explored to design new quantum algorithms.

\subsection{Sampling algorithm}

The Hamiltonian is specified by a vector of real numbers $\bfh = (h_1,\ldots,h_M)$ and a vector of Pauli operators $\bfsigma = (\sigma_1,\ldots,\sigma_M)$. Given the evolution time $t$, we need to choose a number of time steps $N$; then, the corresponding time step size is $\Delta t = t/N$. These parameters, $\bfh$, $\bfsigma$, $N$, and $\Delta t$, are inputs to the sampling algorithm. To present the algorithm in a way that works for both POE and LOR formulas, we introduce an additional input parameter ${\rm F} = {\rm P},{\rm R}$ to denote POE and LOR formulas, respectively.

\begin{algorithm}[H]
{\small
\begin{algorithmic}[1]
\caption{{\small Sample generation.}}
\label{alg:SamGen}

\Statex
\Function{SamGen}{${\rm F},\bfh,\bfsigma,N,\Delta t$}
\For{$i=1$ to $N$}
\State $(W_i,\theta_i) \leftarrow$ \Call{SamGenOneStep}{${\rm F},\bfh,\bfsigma,\Delta t$}
\State $(W_i',\theta_i') \leftarrow$ \Call{SamGenOneStep}{${\rm F},\bfh,\bfsigma,\Delta t$}
\EndFor
\State $\boldsymbol{W} \leftarrow (W_1,\ldots,W_N,W_1',\ldots,W_N')$
\State $\theta \leftarrow \sum_{i=1}^N \left( \theta_i - \theta_i' \right)$
\State Output $(\boldsymbol{W},\theta)$.
\EndFunction
\end{algorithmic}
}
\end{algorithm}

\begin{algorithm}[H]
{\small
\begin{algorithmic}[1]
\caption{{\small Sample generation for one time step.}}
\label{alg:OneStep}

\Statex
\Function{SamGenOneStep}{${\rm F},\bfh,\bfsigma,\Delta t$}

\State Compute $L_2$ according to Eq.~(\ref{eq:L2}), simplify $L_2$ by combining like terms, obtain the final expression $L_2 = \sum_u \alpha_u \tau_u$.
\Comment $\tau_u\in \bfP_n$

\State $C_L \leftarrow \sum_u \abs{\alpha_u}$

\State $C_T \leftarrow e^{2h_{\rm tot}\Delta t} - \sum_{k=0}^{5}\frac{1}{k!}(2h_{\rm tot}\Delta t)^k$
\Comment $h_{\rm tot} = \sum_j \abs{h_j}$

\If {${\rm F} = {\rm P}$}
$C_A \leftarrow 1+C_L+C_T$
\ElsIf {${\rm F} = {\rm R}$}
$C_A \leftarrow \sqrt{1+C_L^2}+C_T$
\EndIf

\State Choose ${\rm O}$ from ${\rm L}$ and ${\rm T}$ with probabilities $(C_A-C_T)/C_A$ and $C_T/C_A$, respectively.

\If {${\rm O} = {\rm L}$}
\Comment Sample from leading-order terms

\If {${\rm F} = {\rm P}$}
Choose $(W,\theta)$ from $(\openone,0)$ and $\{(\tau_u,\arg(-i\alpha_u))\}$ with probabilities $1/(1+C_L)$ and $\{\abs{\alpha_u}/(1+C_L)\}$, respectively.
\ElsIf {${\rm F} = {\rm R}$}
Choose $(W,\theta)$ from $\{(e^{-i {\rm sgn}(\alpha_u) \phi \tau_u},0)\}$ with probabilities $\{\abs{\alpha_u}/C_L\}$.
\EndIf
\Comment $\phi = \arctan(C_L)$

\ElsIf {${\rm O} = {\rm T}$}
\Comment Sample from high-order terms

\State $k,k_j,k_j' \leftarrow 0$
\While {$k+\sum_{j=1}^M (k_j+k_j')<6$}
\State $k \leftarrow$ \Call{Poisson}{$h_{\rm tot}\Delta t$}
\Comment \Call{Poisson}{$x$} returns $k\in \{0,1,\ldots\}$ with the probability $e^{-x}x^k/k!$.
\For{$j=1$ to $M$}
\State $k_j \leftarrow$ \Call{Poisson}{$\abs{h_j}\Delta t/2$}
\State $k_j' \leftarrow$ \Call{Poisson}{$\abs{h_j'}\Delta t/2$}
\EndFor
\EndWhile

\For{$a=1$ to $k$} Choose $j_a$ from $\{1,\ldots,M\}$ with probabilities $\{\abs{h_{j_a}}/h_{\rm tot}\}$.
\EndFor

\State $W \leftarrow \zeta^* \sigma_M^{k_M'}\cdots \sigma_1^{k_1'} \sigma_{j_k}\cdots \sigma_{j_1} \sigma_1^{k_1}\cdots \sigma_M^{k_M}$

\Comment Take $\zeta = \pm 1,\pm i$ to meet $W\in \bfP_n$.

\State $\theta \leftarrow \arg\left( \zeta \prod_{a=1}^k \left(-ih_{j_a}\right) \times \prod_{j=1}^M (ih_{j})^{k_j+k_j'} \right)$

\EndIf

\State Output $(W,\theta)$.

\EndFunction
\end{algorithmic}
}
\end{algorithm}

In the second-order summation formulas, each term is in the form $S_{1}(-\frac{\Delta t}{2})^\dag W S_{1}(\frac{\Delta t}{2})$: In the POE formula, $W$ is always a Pauli operator; in the LOR formula, $W$ is either a rotation operator or a Pauli operator. Taking $U(s_i) = S_{1}(-\frac{\Delta t}{2})^\dag W_i S_{1}(\frac{\Delta t}{2})$ and $U(s_i') = S_{1}(-\frac{\Delta t}{2})^\dag W_i' S_{1}(\frac{\Delta t}{2})$, we have
\begin{eqnarray}
O_{\bfs} &=& S_1^\dag W_1^{\prime\dag} S_1^{\prime\dag} \cdots S_1^\dag W_N^{\prime\dag} S_1^{\prime\dag} \notag \\
&&\times O S_1' W_N S_1 \cdots S_1' W_1 S_1,
\label{eq:Oss}
\end{eqnarray}
Here, we use the notations $S_1 = S_1(\frac{\Delta t}{2})$ and $S_1' = S_1(-\frac{\Delta t}{2})^\dag$ for simplicity. Given the vector of correction operators
\begin{eqnarray}
\boldsymbol{W} = (W_1,\ldots,W_N,W_1',\ldots,W_N'),
\label{eq:W}
\end{eqnarray}
the quantum computer can evaluate $\bra{\psi_{\rm f}} O_{\bfs} \ket{\psi_{\rm i}}$.

In the final phase, the classical computer estimates the transition amplitude by computing the expected value of $e^{i\theta_{\bfs}}\bra{\psi_{\rm f}} O_{\bfs} \ket{\psi_{\rm i}}$. Therefore, the sampling algorithm also needs to output $\theta_{\bfs}$.

Overall, the outputs of the sampling algorithm are $\boldsymbol{W}$ and $\theta$. The procedure for generating $\boldsymbol{W}$ and $\theta$ is given in Algorithm~\ref{alg:SamGen}. Algorithm~\ref{alg:OneStep} is a subroutine for processing one time step.

\subsection{Quantum-circuit Monte Carlo algorithm}

Using the Monte Carlo summation to compute the path-integral formula in Eq.~(\ref{eq:QCMC}), we need to choose two parameters $N_{\rm s}$ and $M_{\rm s}$, which are the number of $\bfs$ samples and the number of shots per quantum circuit for evaluating $\bra{\psi_{\rm f}} O_{\bfs} \ket{\psi_{\rm i}}$, respectively. Because the transition amplitude is a complex number in general, the quantum computing returns two real numbers $a_{{\rm R},\bfs}$ and $a_{{\rm I},\bfs}$, which are estimates of the real and imaginary parts of $e^{i\theta_{\bfs}}\bra{\psi_{\rm f}} O_{\bfs} \ket{\psi_{\rm i}}$, respectively. By computing expected values of $a_{{\rm R},\bfs}$ and $a_{{\rm I},\bfs}$, we obtain the transition amplitude $\bra{\psi_{\rm f}} e^{iHt} O e^{-iHt} \ket{\psi_{\rm i}}$ up to the factor $C_A^{2N}$. QCMC is summarised in Algorithm~\ref{alg:QCMC}.

\begin{algorithm}[H]
{\small
\begin{algorithmic}[1]
\caption{{\small Quantum-circuit Monte Carlo.}}
\label{alg:QCMC}

\Statex
\State Input ${\rm F},\bfh,\bfsigma,N,\Delta t,N_{\rm s},M_{\rm s}$.
\For{$v=1$ to $N_{\rm s}$}
\Comment $v$ is the label of $\bfs$ samples.
\State $(\boldsymbol{W},\theta) \leftarrow$ \Call{SamGen}{${\rm F},\bfh,\bfsigma,N,\Delta t$}
\State $(a_{{\rm R},v},a_{{\rm I},v}) \leftarrow$ \Call{QuantumCircuits}{$\ldots,\boldsymbol{W},\theta,M_{\rm s}$}
\EndFor
\State $\hat{A} \leftarrow C_A^{2N} \frac{1}{N_{\rm s}}\sum_{v=1}^{N_{\rm s}}\left(a_{{\rm R},v} + ia_{{\rm I},v}\right)$
\State Output $\hat{A}$ as the estimate of $\bra{\psi_{\rm f}} e^{iHt} O e^{-iHt} \ket{\psi_{\rm i}}$.
\end{algorithmic}
}
\end{algorithm}

\subsection{Quantum-circuit Monte Carlo on classical computer}
\label{sec:CC}

In this section, we show that QCMC with the zeroth-order POE formula is equivalent to QMC on a classical computer. In the zeroth-order POE formula, the time evolution operator is expanded into the form $e^{-iH\Delta t} = \sum_s c(s) \sigma_s$, where the $\sigma_s\in \bfP_n$ are Pauli operators. We can express a Pauli operator as
\begin{eqnarray}
\sigma = i^{x_1z_1}X^{x_1}Z^{z_1}\otimes\cdots\otimes i^{x_nz_n}X^{x_n}Z^{z_n},
\label{eq:sigma}
\end{eqnarray}
where $x_a,z_a = 0,1$, and $i^{x_az_a}X^{x_a}Z^{z_a} = I,X,Y,Z$ is a single-qubit Pauli operator of qubit-$a$. We consider computational basis states in the form $\bigotimes_{a=1}^n \ket{\mu_a}$, where $\mu_a = 0,1$. A Pauli operator acting on a basis state always results in a basis state, i.e.~$\sigma \bigotimes_{a=1}^n \ket{\mu_a} = \bigotimes_{a=1}^n i^{x_az_a} (-1)^{z_a\mu_a} \ket{\mu_a\oplus x_a}$, where $\oplus$ denotes the modulo 2 addition. Therefore, Pauli operators acting on basis states can be efficiently calculated on a classical computer. Similarly, Pauli operators acting on product states in the form $\bigotimes_{j=1}^n \ket{\psi_j}$ and stabiliser states~\cite{Gottesman1998} can also be efficiently calculated on a classical computer. In the following, we focus on computational basis states.

The zeroth-order POE formula is auxiliary-field Monte Carlo, which takes the space of Pauli operators as the auxiliary-field. Suppose that the initial and final states are computational basis states and $O$ is a Pauli operator. We can evaluate
\begin{eqnarray}
\bra{\psi_{\rm f}} O_{\bfs} \ket{\psi_{\rm i}} = \bra{\psi_{\rm f}} \sigma_{s_1'} \cdots \sigma_{s_N'} O \sigma_{s_N} \cdots \sigma_{s_1} \ket{\psi_{\rm i}}
\end{eqnarray}
on a classical computer. By expressing the initial and final states as linear combinations of basis states and the operator $O$ as a linear combination of Pauli operators, we can evaluate $\bra{\psi_{\rm f}} O_{\bfs} \ket{\psi_{\rm i}}$ on a classical computer for the general states and the operator. Therefore, we can implement QCMC with the zeroth-order POE formula without using a quantum computer.

Now, we consider a class of Hamiltonians without short-time interference between Pauli operators. Each Pauli operator corresponds to two binary strings $(x_1,\ldots,x_n)$ and $(z_1,\ldots,z_n)$. If the $x$ strings of two Pauli operators $\sigma$ and $\tau$ are different, we have $\bra{\psi} \tau \sigma \ket{\psi} = 0$ for all computational basis states $\ket{\psi} = \bigotimes_{a=1}^n \ket{\mu_a}$. The short-time evolution operator $e^{-iH\Delta t} \simeq \openone - iH\Delta t$ acting on a basis state results in
\begin{eqnarray}
e^{-iH\Delta t} \ket{\psi} \simeq \ket{\psi} - i\Delta t\sum_j h_j \sigma_j \ket{\psi}.
\end{eqnarray}
We find that there is no interference between the terms if and only if $\bra{\psi} \tau \sigma \ket{\psi} = 0$ for all $\sigma,\tau \in \{\openone\}\cup\{\sigma_j\}$: i.e., the Pauli operators in the Hamiltonian have different $x$ strings.

For Hamiltonians without short-time interference, the zeroth-order POE formula is equivalent to Green's function Monte Carlo, which takes the computational basis. In Green's function Monte Carlo, we sample states $\ket{\bfr}$; in QCMC, we sample Pauli operators. Substituting the computational basis for $\{\ket{\bfr}\}$, the transition amplitude of each time step reads $\bra{\psi'}e^{-iH\Delta t}\ket{\psi}$, where $\ket{\psi'} = \bigotimes_{a=1}^n \ket{\mu_a'}$. For a Hamiltonian without short-time interference, basis states $\ket{\psi'}$ with nonzero $\bra{\psi'}e^{-iH\Delta t}\ket{\psi}$ and Pauli operators in $\{\openone\}\cup\{\sigma_i\}$ have one-to-one correspondence in the limit of small $\Delta t$. Therefore, sampling Pauli operators is equivalent to sampling basis states $\ket{\psi'}$.

The class of Hamiltonians without short-time interference includes those are hard for simulation in classical computing. In Appendix~\ref{app:FHM}, we show that the Fermi-Hubbard model on any bipartite lattice (e.g.~the square lattice) can be encoded into a qubit Hamiltonian without short-time interference, using the Jordan-Wigner transformation.


\section{Quantum circuits}
\label{sec:circuits}

We propose quantum circuits for evaluating the transition amplitude of the operator $O_{\bfs}$, and the gate number per time step is moderately increased upon the Lie-Trotter-Suzuki product. To measure the transition amplitude, we need to introduce an ancillary qubit, which controls the evolution of $n$ qubits representing the system. In Eq.~(\ref{eq:Oss}), the evolution is driven by Lie-Trotter-Suzuki products $R_1$ and correction operators $W$. Our circuits are simplified in two ways: first, we avoid controlled Lie-Trotter-Suzuki products and only use controlled corrections; and, second, the correction operators are either Pauli operators $\sigma$ or rotation operators $e^{-i\phi\sigma}$.

We propose two types of circuits. For compact circuits, the circuit depth is the same as the Lie-Trotter-Suzuki decomposition with additional controlled-correction gates. For forward-backward circuits, the circuit depth is doubled, but they provide inherent quantum error mitigation. In this section, we also show how to efficiently decompose a controlled-correction gate into elementary gates. We assume that $O$ is a unitary operator, and we can evaluate a general operator by decomposing it into a linear combination of unitary operators.

\begin{figure*}[tbp]
\begin{center}
\includegraphics[width=1\linewidth]{\figpath/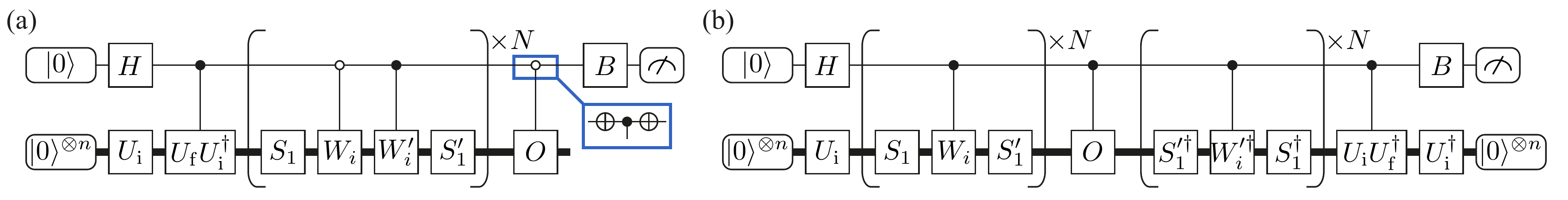}
\caption{
Quantum circuits for evaluating $e^{i\theta_{\bfs}}\bra{\psi_{\rm f}} O_{\bfs} \ket{\psi_{\rm i}}$. The qubit on the top is the ancillary qubit. The empty circle in the blue box denotes a controlled-$U$ gate that $U$ acts on the $n$ qubits when the ancillary qubit is in $\ket{0}$. Unitary operators $U_{\rm i}$ and $U_{\rm f}$ prepare the initial and final states, respectively, i.e.~$\ket{\psi_{\rm i}} = U_{\rm i} \ket{0}^{\otimes n}$ and $\ket{\psi_{\rm f}} = U_{\rm f} \ket{0}^{\otimes n}$. The gate $B$ is for adjusting the measurement basis. For simplicity, we use the notations $S_1 = S_1(\frac{\Delta t}{2})$ and $S_1' = S_1(-\frac{\Delta t}{2})^\dag$.
}
\label{fig:circuit}
\end{center}
\end{figure*}

\subsection{Compact circuit}
\label{sec:compact}

The compact circuit for second-order formulas is shown in Fig.~\ref{fig:circuit}(a). The circuits for the other summation formulas are similar. For the first-order formulas, we remove the $S_1'$ products from the circuit; for the zeroth-order formulas, we remove both the $S_1$ and the $S_1'$ products; and by adding more $S_1$ and $S_1'$ products, the circuit can be used for higher-order formulas. If we ignore controlled-correction gates, the compact circuit for the $l$th-order summation formula is the same as the circuit for the $l$th-order Lie-Trotter-Suzuki product formula.

Now, we focus on second-order formulas, and the analysis for the other formulas is similar. The final state of the compact circuit (before the basis adjusting gate $B$) is
\begin{eqnarray}
\ket{\Psi} &=& \frac{1}{\sqrt{2}} \left( \ket{0}_{\rm a}\otimes O S_1' W_N S_1 \cdots S_1' W_1 S_1 \ket{\psi_{\rm i}} \right. \notag \\
&&+ \left. \ket{1}_{\rm a}\otimes S_1' W_N' S_1 \cdots S_1' W_1' S_1 \ket{\psi_{\rm f}} \right).
\end{eqnarray}
Measuring the ancillary qubit, we obtain
\begin{subequations}
\begin{eqnarray}
\bra{\Psi}X_a\ket{\Psi} &=& \Re \left( \bra{\psi_{\rm f}} O_{\bfs} \ket{\psi_{\rm i}} \right),
\end{eqnarray}
\begin{eqnarray}
\bra{\Psi}Y_a\ket{\Psi} &=& -\Im \left( \bra{\psi_{\rm f}} O_{\bfs} \ket{\psi_{\rm i}} \right),
\end{eqnarray}
\end{subequations}
where $X_a$ and $Y_a$ are Pauli operators of the ancillary qubit. Here, we use Eq.~(\ref{eq:Oss}). The procedure for evaluating $\bra{\psi_{\rm f}} O_{\bfs} \ket{\psi_{\rm i}}$ using compact circuits is given in Algorithm~\ref{alg:QC}.

\begin{algorithm}[H]
{\small
\begin{algorithmic}[1]
\caption{{\small Quantum circuit evaluation.}}
\label{alg:QC}

\Statex
\Function{QuantumCircuits}{$\bfh,\bfsigma,N,\Delta t,\boldsymbol{W},\theta,M_{\rm s}$}
\State $a_{\rm R},a_{\rm I} \leftarrow 0$
\For{$i=1$ to $M_{\rm s}$}
\State Implement the circuit, measure $\cos\theta X_a + \sin\theta Y_a$ and collect the outcome $\mu_{\rm R} = \pm 1$.
\State $a_{\rm R} \leftarrow a_{\rm R} + \mu_{\rm R}$
\State Implement the circuit, measure $\sin\theta X_a - \cos\theta Y_a$ and collect the outcome $\mu_{\rm I} = \pm 1$.
\State $a_{\rm I} \leftarrow a_{\rm I} + \mu_{\rm I}$
\EndFor
\State $a_{\rm R} \leftarrow a_{\rm R}/M_{\rm s}$
\State $a_{\rm I} \leftarrow a_{\rm I}/M_{\rm s}$
\State Output $(a_{\rm R},a_{\rm I})$.
\EndFunction
\end{algorithmic}
}
\end{algorithm}

\subsection{Forward-backward circuit}
\label{sec:FBcircuit}

The forward-backward circuit for second-order formulas is shown in Fig.~\ref{fig:circuit}(b). Compared with the compact circuit, the number of $S_1$ and $S_1'$ products is doubled. The final state of the circuit is
\begin{eqnarray}
\ket{\Phi} &=& \frac{1}{\sqrt{2}} \left( \ket{0}_{\rm a}\otimes \ket{0}^{\otimes n} + \ket{1}_{\rm a}\otimes U_{\rm f}^\dag O_{\bfs} \ket{\psi_{\rm i}} \right).
\label{eq:Phi}
\end{eqnarray}
Here, we use Eq.~(\ref{eq:Oss}). Measuring the ancillary qubit, we obtain
\begin{subequations}
\begin{eqnarray}
\bra{\Phi}X_a\ket{\Phi} &=& \Re \left( \bra{\psi_{\rm f}} O_{\bfs} \ket{\psi_{\rm i}} \right),
\end{eqnarray}
\begin{eqnarray}
\bra{\Phi}Y_a\ket{\Phi} &=& \Im \left( \bra{\psi_{\rm f}} O_{\bfs} \ket{\psi_{\rm i}} \right).
\end{eqnarray}
\label{eq:FBcircuit}
\end{subequations}
The procedure for evaluating $\bra{\psi_{\rm f}} O_{\bfs} \ket{\psi_{\rm i}}$ is similar to Algorithm~\ref{alg:QC}. Note that evaluating the transition amplitude in this way does not provide inherent error mitigation. We discuss the inherent error mitigation using postselection in Sec.~\ref{sec:inherent}.

\subsection{Controlled-correction gates}
\label{sec:correction}

\begin{figure}[tbp]
\begin{center}
\includegraphics[width=1\linewidth]{\figpath/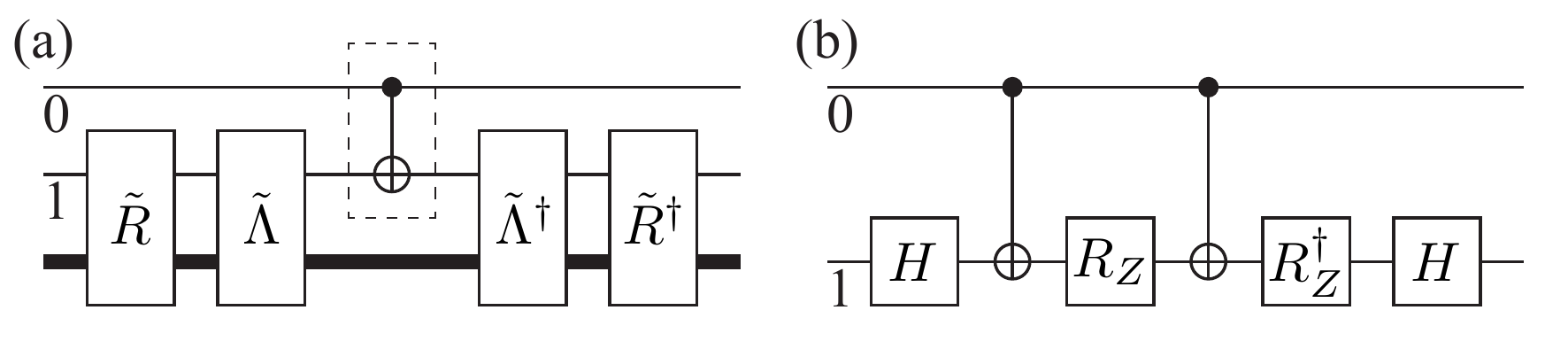}
\caption{
(a) The circuit of the controlled-$\sigma$ gate. Qubit $0$ is the ancillary qubit. (b) The circuit of the controlled-$e^{-i\phi X_1}$ gate. Replacing the controlled-NOT gate in the dashed box with the controlled-$e^{-i\phi X_1}$ gate, we have the circuit of the controlled-$e^{-i\phi\sigma}$ gate. The single-qubit phase gate $R_Z = e^{i\frac{\phi}{2}Z}$.
}
\label{fig:circuitII}
\end{center}
\end{figure}

We consider two types of qubit networks. On the all-to-all network, controlled-NOT gates on all pairs of qubits are available. On the linear network, only controlled-NOT gates on nearest neighboring qubits are allowed. We use $\Lambda_{a,b}$ to denote the controlled-NOT gate that $a$ and $b$ are the control and target qubits, respectively. Because the error rate of controlled-NOT gates is usually much higher than that of single-qubit gates, we only count controlled-NOT gates and minimise their number.

A general Pauli operator $\sigma$ is equivalent to an $X$-product Pauli operator (i.e.~a tensor product of $I$ and $X$) up to a unitary transformation. For the $\sigma$ in Eq.~(\ref{eq:sigma}), the transformation is $\tilde{R} = H^{z_1(1-x_1)}S^{z_1x_1}\otimes\cdots\otimes H^{z_n(1-x_n)}S^{z_nx_n}$, where $H$ is the Hadamard gate, and $S$ is the $\frac{\pi}{4}$ phase gate. This transformation leads to $\tilde{\sigma} = \tilde{R}^\dag \sigma \tilde{R} = X^{x_1\vee z_1}\otimes\cdots\otimes X^{x_n\vee z_n}$, where $x_a\vee z_a = 1-(1-x_a)(1-z_a)$.

Implementation of the controlled-$\tilde{\sigma}$ gate on the all-to-all network is straightforward. For each qubit with $x_a\vee z_a = 1$, we apply the controlled-NOT gate $\Lambda_{0,a}$, where qubit $0$ is the ancillary qubit. The controlled-$\tilde{\sigma}$ gate is $\prod_{a=1}^n \Lambda_{0,a}^{x_a\vee z_a}$, and the number of controlled-NOT gates is $N_\Lambda = \sum_{a=1}^{n} x_a\vee z_a \leq n$.

In the compact circuit shown in Fig.~\ref{fig:circuit}(a), there are two controlled-correction gates in each time step, corresponding to $W_i$ and $W_i'$, respectively. When $W_i = \tau$ and $W_i' = \tau'$ are Pauli operators, we can combine the two controlled-correction gates into one controlled-$\sigma$ gate in the following way. Note that $\tau'\tau = \zeta \sigma$, where $\zeta$ is a phase factor. We apply $\tau$ first, then a controlled-$\sigma$ gate, and finally a phase gate $\mathrm{diag}(1,\zeta)$ on the ancillary qubit. The overall transformation is equivalent to the two controlled-correction gates. The total number of controlled-NOT gates is $N_\Lambda$.

Now, we present another protocol for the controlled-$\sigma$ gate. The circuit is shown in Fig.~\ref{fig:circuitII}(a), which is formed of three parts: gates transforming a general Pauli operator $\sigma$ into an $X$-product Pauli operator $\tilde{\sigma}$, gates transforming $\tilde{\sigma}$ into the single-qubit Pauli operator $X_1$ on qubit $1$, and the controlled-NOT gate $\Lambda_{0,1}$ on the ancillary qubit and qubit $1$. On the linear network, we assume that qubit $1$ is next to the ancillary qubit. On the all-to-all network, we can label any qubit as qubit $1$; without loss of generality, we assume that $x_1\vee z_1 = 1$. In this protocol, there is only one instead of $N_\Lambda$ gates on the ancillary qubit. Because the outcome is obtained by measuring the ancillary qubit, applying fewer gates on the ancillary qubit potentially reduces the impact of errors. Replacing $\Lambda_{0,1}$ with the circuit in Fig.~\ref{fig:circuitII}(b), we can realise the controlled-$e^{-i\phi\sigma}$ gate.

To transform $\tilde{\sigma}$ into $X_1$, we look for a transformation $\tilde{\Lambda}$ that satisfies $\sigma = \tilde{\Lambda}^\dag X_1 \tilde{\Lambda}$. On the all-to-all network, we take $\tilde{\Lambda} = \prod_{a=2}^n \Lambda_{1,a}^{x_a\vee z_a}$ and the number of controlled-NOT gates for each $\tilde{\Lambda}$ is $N_\Lambda - 1$. On the linear network, we take
\begin{eqnarray}
\tilde{\Lambda} = \Lambda_{1,2} \Lambda_{2,1}^{x_1\downarrow z_1} \Lambda_{2,3} \Lambda_{3,2}^{x_2\downarrow z_2} \cdots \Lambda_{n'-1,n'} \Lambda_{n',n'-1}^{x_{n'-1}\downarrow z_{n'-1}},~~
\end{eqnarray}
where $x_a\downarrow z_a = (1-x_a)(1-z_a)$ and $n' = \max\{a\st x_a=1\}$. On the linear network, the number of controlled-NOT gates for each $\tilde{\Lambda}$ is $(n'-1)+\sum_{a=1}^{n'-1} x_a\downarrow z_a \leq 2n-2$.

The maximum number of controlled-NOT gates for implementing the two controlled-correction gates in each time step is summarised as follows. On the all-to-all network, the maximum gate number is $n$ for POE formulas, which becomes $2(n-1)+1 = 2n-1$ to reduce gates on the ancillary qubit and $2[2(n-1)+2] = 4n$ for LOR formulas. On the linear network, the maximum gate number is $2(2n-2)+1 = 4n-3$ for POE formulas and $2[2(2n-2)+2] = 8n-4$ for LOR formulas. In the MCQC algorithm, the controlled-correction gates are randomly selected and the gate number could be much smaller than its maximum value. For example, for the POE formula, the Pauli operator of the zeroth-order term in the expansion is the identity.

\section{Optimal distribution}
\label{sec:distribution}

In this section, we derive the optimal distribution of $\bfs$ that minimises the variance in Monte Carlo. Using the protocols in Sec.~\ref{sec:circuits} to evaluate $\bra{\psi_{\rm f}} O_{\bfs} \ket{\psi_{\rm i}}$, we prove that taking the distribution in Eq.~(\ref{eq:pro}) and $M_{\rm s} = 1$ is optimal and that the minimum variance is given by Eq.~(\ref{eq:var}).

A quantum circuit usually has random measurement outcomes; therefore, the outputs of quantum computing $a_{{\rm R},\bfs}$ and $a_{{\rm I},\bfs}$ are random variables. We suppose that $a_{\nu,\bfs}$ ($\nu = {\rm R},{\rm I}$) takes the value $a_i$ with the probability $P_{\nu,\bfs,i}$ in the quantum computing; then, its expected value is $\mathrm{E}\left[a_{\nu,\bfs}\right]_{\rm qc} = \sum_i P_{\nu,\bfs,i} a_i$. Here, $\mathrm{E}\left[\bullet\right]_{\rm qc}$ denotes the mean taken over quantum computing runs for the specific $\bfs$ (each run returns an output evaluated using $M_{\rm s}$ shots) and $\mathrm{E}\left[\bullet\right]$ without the subscript `QC' denotes the mean taken over both $\bfs$ and quantum computing runs. Using the protocols in Sec.~\ref{sec:circuits}, $a_{{\rm R},\bfs}$ and $a_{{\rm I},\bfs}$ are unbiased estimators of $\bra{\psi_{\rm f}} O_{\bfs} \ket{\psi_{\rm i}}$, i.e.
\begin{eqnarray}
e^{i\theta_{\bfs}}\bra{\psi_{\rm f}} O_{\bfs} \ket{\psi_{\rm i}} = \mathrm{E}\left[a_{{\rm R},\bfs}\right]_{\rm QC} + i\mathrm{E}\left[a_{{\rm I},\bfs}\right]_{\rm QC}.
\end{eqnarray}

Let $A_{\rm R}$ and $A_{\rm I}$ be the real and imaginary parts of $\bra{\psi_{\rm f}} e^{iHt} O e^{-iHt} \ket{\psi_{\rm i}}$, respectively. In the QCMC algorithm, we evaluate the summation $A_\nu = \sum_{\bfs} c_{\bfs} \mathrm{E}\left[a_{\nu,\bfs}\right]_{\rm QC}$ using the Monte Carlo method, where $c_{\bfs} = \absLR{\prod_{i=1}^N c(s_i)c(s_i')^*}$. Given any probability distribution $P(\bfs)$, we have
\begin{eqnarray}
A_\nu = \sum_{\bfs} P(\bfs) \frac{c_{\bfs} \mathrm{E}\left[a_{\nu,\bfs}\right]_{\rm QC}}{P(\bfs)} = \mathrm{E}\left[ \frac{c_{\bfs} a_{\nu,\bfs}}{P(\bfs)} \right].
\end{eqnarray}
Therefore, we can estimate $A_\nu$ by sampling $\bfs$ according to the distribution $P(\bfs)$ and compute the empirical mean of $c_{\bfs} a_{\nu,\bfs} / P(\bfs)$. The variance of the estimator $\hat{A}_\nu$ with $N_{\rm s}$ samples is
\begin{eqnarray}
\mathrm{Var}\left(\hat{A}_\nu\right) &=& \frac{1}{N_{\rm s}} \mathrm{Var}\left(\frac{c_{\bfs} a_{\nu,\bfs}}{P(\bfs)}\right) \notag \\
&=& \frac{1}{N_{\rm s}} \sum_{\bfs} \frac{c_{\bfs}^2 \alpha_{\nu,\bfs}^2}{P(\bfs)} - \frac{A_\nu^2}{N_{\rm s}},
\end{eqnarray}
where $\alpha_{\nu,\bfs} = \sqrt{\mathrm{E}\left[a_{\nu,\bfs}^2\right]_{\rm QC}}$. The optimal distribution that minimises the variance is $P(\bfs) \propto \abs{c_{\bfs}} \alpha_{\nu,\bfs}$, and the minimum variance is
\begin{eqnarray}
\mathrm{Var}\left(\hat{A}_\nu\right) = \frac{1}{N_{\rm s}} \left(\sum_{\bfs} \abs{c_{\bfs}} \alpha_{\nu,\bfs}\right)^2 - \frac{A_\nu^2}{N_{\rm s}}.
\label{eq:varII}
\end{eqnarray}

Now, we consider that $a_{\nu,\bfs}$ is obtained by taking the empirical mean of $M_{\rm s}$ binary numbers. Each binary number takes $\pm 1$ corresponding to the measurement outcome of the ancillary qubit (see Algorithm~\ref{alg:QC}). Then, $a_{\nu,\bfs}$ follows the binomial distribution and
\begin{eqnarray}
\alpha_{\nu,\bfs} = \sqrt{\frac{1+(M_{\rm s}-1)\mathrm{E}\left[a_{\nu,\bfs}\right]_{\rm QC}^2}{M_{\rm s}} }.
\end{eqnarray}
Let $M_{\rm tot}$ be the total number of circuit shots; we have $N_{\rm s}  = M_{\rm tot}/M_{\rm s}$. Substituting $\alpha_{\nu,\bfs}$ and $N_{\rm s}$ into Eq.~(\ref{eq:varII}), we obtain the variance as a function of $M_{\rm s}$. Taking $M_{\rm s}$ as a continuous variable, we find that the derivative of the variance with respect to $M_{\rm s}$ is always positive when $M_{\rm s}\geq 1$. Therefore, the variance is minimised at $M_{\rm s} = 1$. When $M_{\rm s} = 1$, we have $\alpha_{\nu,\bfs} = 1$, and the optimal distribution is $P(\bfs) \propto \abs{c_{\bfs}}$. Accordingly, the minimum variance is
\begin{eqnarray}
\mathrm{Var}\left(\hat{A}_\nu\right) = \frac{1}{M_{\rm tot}} \left(\sum_{\bfs} \abs{c_{\bfs}}\right)^2 - \frac{A_\nu^2}{M_{\rm tot}}.
\end{eqnarray}
With $\mathrm{Var}\left(\hat{A}\right) = \mathrm{Var}\left(\hat{A}_{\rm R}\right) + \mathrm{Var}\left(\hat{A}_{\rm I}\right)$, we obtain the minimum total variance in Eq.~(\ref{eq:var}). Here, we assume that the total number of shots for each of the real and imaginary parts is $M_{\rm tot}$.

We remark that the optimal distribution is obtained by assuming the empirical mean estimator for $a_{\nu,\bfs}$. If we have prior knowledge of the $a_{\nu,\bfs}$ distribution, we can use other estimators such as the Bayes estimator to reduce the variance. In the extreme case, suppose that $\mathrm{E}\left[a_{\nu,\bfs}\right]_{\rm QC}$ is known, the optimal distribution is $P(\bfs) \propto \abs{c_{\bfs} \mathrm{E}\left[a_{\nu,\bfs}\right]_{\rm QC}}$ instead of $P(\bfs) \propto \abs{c_{\bfs}} \alpha_{\nu,\bfs}$ (note that in this case, we do not even need the quantum computer).

\section{Quantum Error mitigation}
\label{sec:qem}

Many quantum error mitigation protocols can be classified into three categories. In the first category, with knowledge of the error model, we compensate the effect of errors by using approaches such as error extrapolation and probabilistic error cancellation (i.e.~quasi-probability decomposition)~\cite{Li2017, Temme2017, Endo2018}. In the second category, data from quantum circuits are processed according to constraints on the quantum state. The protocols in this category include, for example, symmetry-based postselection~\cite{McArdle2019PRL, Bonet2018} and purification~\cite{Koczor2020, Huggins2020, Czarnik2021}. There are also protocols, e.g.~subspace expansion~\cite{McClean2017}, introduced for specific algorithms, which belong to the third category.

In this section, we first discuss the application of quasi-probability decomposition in QCMC, and then we show that the forward-backward circuit in Fig.~\ref{fig:circuit}(b) provides inherent error mitigation based on constraints on the state. The error mitigation increases the variance in Monte Carlo. On a noisy quantum computer, we need to choose an optimal time step size $\Delta t$ to minimise the variance. Eventually, the variance is determined by the error rate, which is discussed in Sec.~\ref{sec:QandC}.

\begin{figure}[tbp]
\begin{center}
\includegraphics[width=1\linewidth]{\figpath/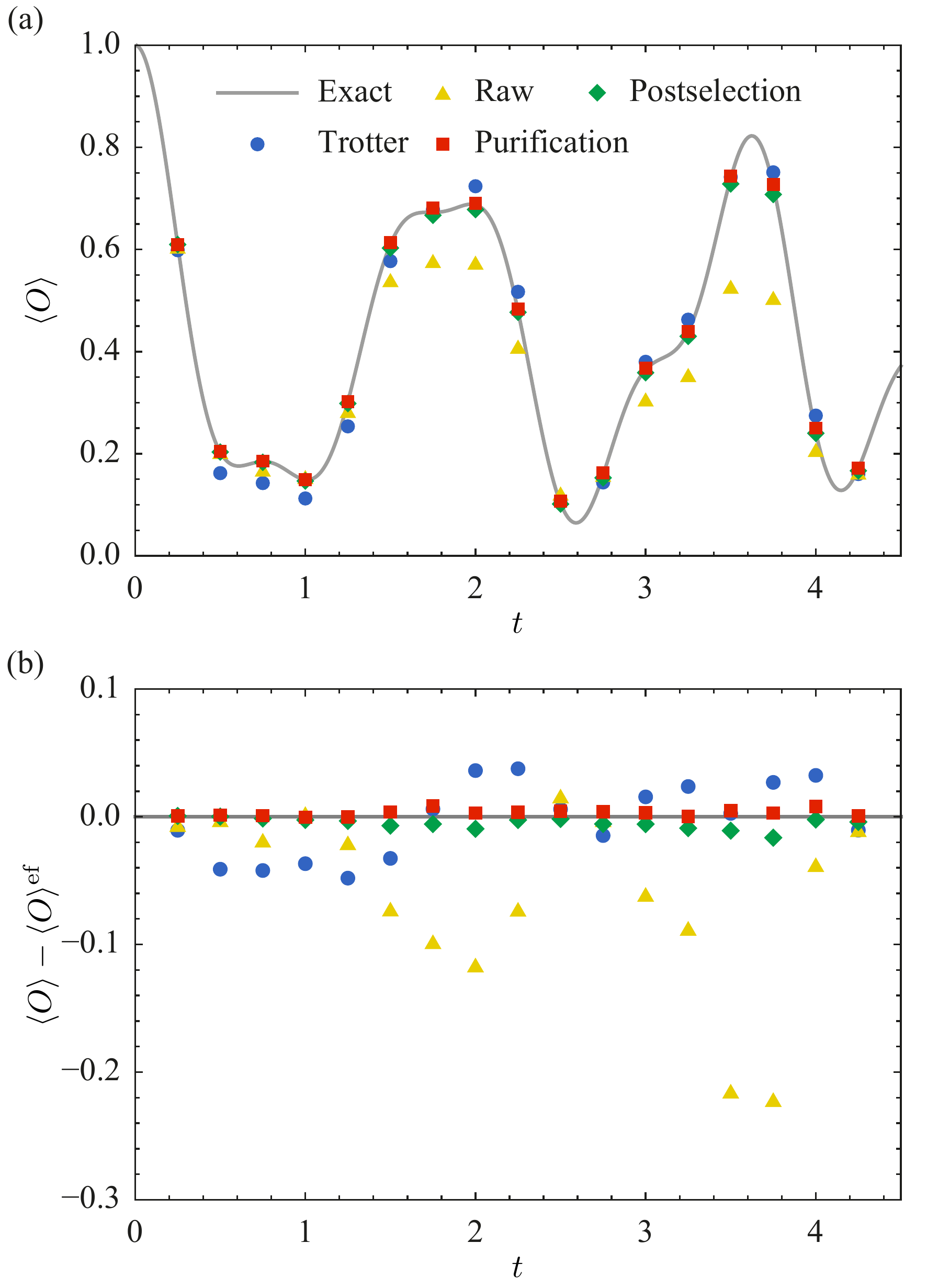}
\caption{
(a)The observable $\mean{O}$ as a function of the time $t$. In the simulation, we take $J=2$, $U=4$ and $\Delta t = 0.05$. The raw data are obtained using the forward-backward circuit without error mitigation, according to Eq.~(\ref{eq:FBcircuit}). With postselection, the observable is computed according to Eq.~(\ref{eq:OssII}). Tomography purification is used to further reduce the error. As a comparison, the Lie-Trotter-Suzuki product formula is evaluated without machine errors. (b) The error $\mean{O} - \mean{O}^{\rm ef}$ in the observable. Here, $\mean{O}^{\rm ef}$ denotes the exact value.
}
\label{fig:FHM}
\end{center}
\end{figure}

\subsection{Quasi-probability decomposition}

In the quasi-probability decomposition, an error-free quantum operation is expressed as a linear combination of noisy operations. Let $\calG^{\rm ef} = [U]$ and $\calG_i$ be the error-free operation and noisy operations, respectively. The quasi-probability decomposition is in the form
\begin{eqnarray}
\calG^{\rm ef} = \sum_i q_i \calG_i,
\end{eqnarray}
where $q_i$ are real coefficients, i.e.~quasi-probabilities. Here, $U$ is a unitary quantum gate, $[U](\bullet) = U\bullet U^\dag$ is the trace-preserving completely positive map of the gate and $\calG_i$ are operations that can actually be implemented on the noisy quantum computer. Similar decompositions can be applied to the initial state and measurement.

We take the Pauli error model as an example. Note that a general error model can be converted into the Pauli error model using Pauli twirling~\cite{Wallman2016}. In the Pauli error model, the noisy operation of a two-qubit gate $U$ reads $\calG = \mathcal{N}[U]$, where the noise map is
\begin{eqnarray}
\mathcal{N} = (1-p)[I\otimes I] + \sum_{\sigma \in \{I,X,Y,Z\}^{\otimes 2}\setminus \{I\otimes I\}} p_\sigma [\sigma],
\label{eq:noise}
\end{eqnarray}
$p_\sigma \ll 1$ is the rate of Pauli error $\sigma$, and $p = \sum_{\sigma\neq I\otimes I} p_\sigma$ is the total error rate. The inverse map of $\mathcal{N}$ is also in the Pauli-operation summation form, i.e.
\begin{eqnarray}
\mathcal{N}^{-1} = \sum_{\sigma \in \{I,X,Y,Z\}^{\otimes 2}} q_\sigma [\sigma],
\end{eqnarray}
and we can solve coefficients $q_\sigma$ numerically. Without a general analytically expression of $q_\sigma$, it is sufficient for us to consider the first-order expansion in order to discuss the impact on variance. To the first order, we have
\begin{eqnarray}
q_{I\otimes I} &=& 1 + p + O(p^2), \\
q_{\sigma\neq I\otimes I} &=& -p_\sigma + O(p^2).
\end{eqnarray}
Given the inverse map, the quasi-probability decomposition of gate $U$ is
\begin{eqnarray}
\calG^{\rm ef} = \sum_{\sigma \in \{I,X,Y,Z\}^{\otimes 2}} q_\sigma [\sigma] \calG.
\end{eqnarray}
Assuming that errors in single-qubit gates are negligible, the composite operation $[\sigma] \calG$ can be implemented on the noisy quantum computer by adding a Pauli gate $\sigma$ after the noisy two-qubit gate $\calG$. We note that the assumption of negligible errors in single-qubit gates is not necessary for the quasi-probability decomposition.

Now, we apply the quasi-probability decomposition to a quantum circuit. In QCMC, using protocols in Sec.~\ref{sec:circuits}, only the ancillary qubit is measured. We can adjust the measurement basis using the gate $B$ in Fig.~\ref{fig:circuit}; therefore, without loss of generality, we focus on the observable $Z_a$ in the error mitigation. Given a quantum circuit formed of many elementary gates, the mean of $Z_a$ reads
\begin{eqnarray}
\mean{Z_a} = \Tr\left[ Z_a \calG_{N_G} \cdots \calG_1 (\rho) \right],
\end{eqnarray}
where $\rho = \ketbra{0}{0}^{\otimes n}$ is the initial state of the quantum circuit, and $N_G$ is the number of gates. Suppose that the quasi-probability decomposition of each gate is $\calG_j^{\rm ef} = \sum_i q_{j,i} \calG_{j,i}$. The error-free mean is
\begin{eqnarray}
\mean{Z_a}^{\rm ef} &=& \sum_{i_1,\ldots,i_{N_G}} \left(\prod_{j=1}^{N_G} q_{j,i}\right) \notag \\
&&\times \Tr\left[ Z_a \calG_{N_G,i_{N_G}} \cdots \calG_{1,i_1} (\rho) \right].
\label{eq:qp}
\end{eqnarray}
Each term in the summation is the mean of $Z_a$ in a circuit modified from the original one. We remark that errors in the initial state and the final measurement can be corrected in a similar way.

We evaluate the decomposition formula in Eq.~(\ref{eq:qp}) using the Monte Carlo summation method by sampling random noisy circuits; therefore, such an error mitigation protocol is called probabilistic error cancellation. Similar to QCMC, the sampling of random circuits increases the variance by a factor of $C_E^2$, where $C_E = \prod_{j=1}^{N_G} \left(\sum_i \abs{q_{j,i}}\right)$. According to the Pauli error model, we have $C_E = \prod_{j=1}^{N_G} [1+2p_j+O(p_j^2)]$, where $p_j$ is the error rate of the $j$th gate. We find that the factor $C_E$ increases with the number of noisy gates. Therefore, the circuit with fewer gates, i.e.~the compact circuit in Fig.~\ref{fig:circuit}(a), is preferred.

In previous discussions, we have assumed that errors in different gates are not correlated. To deal with correlations, we need to introduce a general form of the quasi-probability decomposition,
\begin{eqnarray}
\mean{Z_a}^{\rm ef}_{\bfC_0} &=& \sum_k q_k \mean{Z_a}_{\bfC_k},
\label{eq:qpII}
\end{eqnarray}
where $\mean{Z_a}_{\bfC_k}$ is the mean of $Z_a$ in the circuit $\bfC_k$, $\bfC_0$ is the original circuit, and the $\bfC_{k\neq 0}$ are modified circuits. Modified circuits generated by adding single-qubit operations to the original circuit are usually sufficient for the existence of the decomposition formula. Without correlations, we can work out quasi-probabilities using gate set tomography~\cite{Endo2018}; with correlations, we can determine quasi-probabilities using data of Clifford circuits, i.e.~Clifford sampling~\cite{Strikis2020, Czarnik2020}. Given the quasi-probability decomposition formulas, we can evaluate $\bra{\psi_{\rm f}} O_{\bfs} \ket{\psi_{\rm i}}$ with error mitigation following the procedure in Algorithm~\ref{alg:QEM}.

\begin{algorithm}[H]
{\small
\begin{algorithmic}[1]
\caption{{\small Quantum circuit evaluation with error mitigation.}}
\label{alg:QEM}

\Statex
Input $\bfh,\bfsigma,N,\Delta t,\boldsymbol{W},\theta,M_{\rm s},\nu$.
\State Compose the circuit $\bfC_0$ according to input parameters, in which $\nu = {\rm R},{\rm I}$ (real or imaginary) determines the basis adjusting gate $B$ in Fig.~\ref{fig:circuit}.
\State Work out the decomposition formula in Eq.~(\ref{eq:qpII}).
\State $C_E \leftarrow \sum_k \abs{q_k}$
\State $a_\nu \leftarrow 0$
\For{$i=1$ to $M_{\rm s}$}
\State Choose $k$ with the probability $\abs{q_k}/C_E$.
\State Implement the circuit $\bfC_k$, measure $Z_a$ and collect the outcome $\mu = \pm 1$.
\State $a_\nu \leftarrow a_\nu + C_E \mu$
\EndFor
\State $a_\nu \leftarrow a_\nu/M_{\rm s}$
\State Output $a_\nu$.
\end{algorithmic}
}
\end{algorithm}

\subsection{Inherent error mitigation by postselection}
\label{sec:inherent}

\begin{figure}[tbp]
\begin{center}
\includegraphics[width=1\linewidth]{\figpath/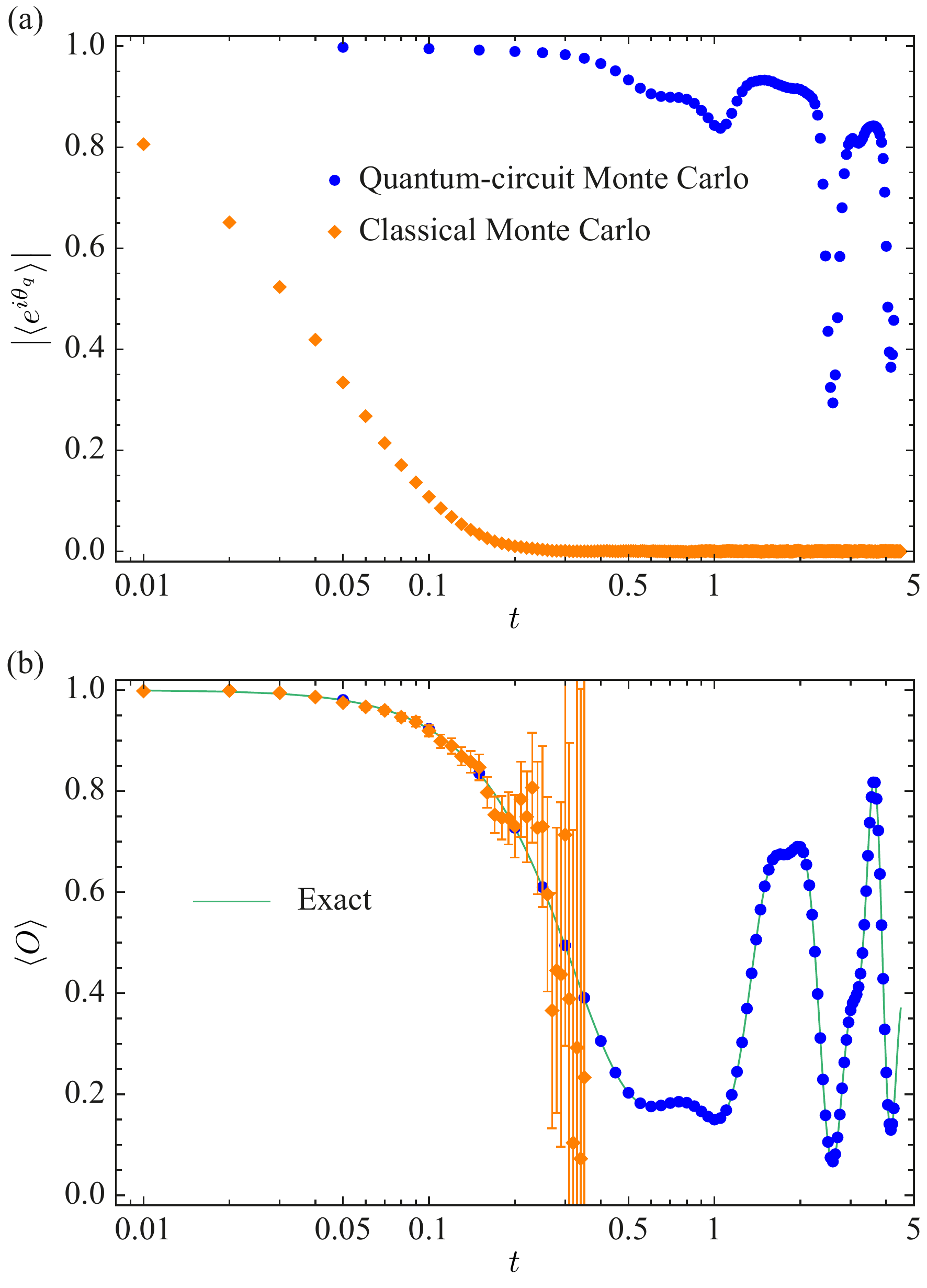}
\caption{
(a) The phase average and (b) the expected value of the observable $O = 2c_{3,\uparrow}^\dag c_{3,\uparrow} - \openone$ in the Monte Carlo simulation of the Fermi-Hubbard model. The Hamiltonian is given in Eq.~(\ref{eq:FHM}), in which we take $J=2$ and $U=4$. The simulation is to compute $\mean{O} = \bra{\psi_{\rm i}}e^{iHt}Oe^{-iHt}\ket{\psi_{\rm i}}$, where $\ket{\psi_{\rm i}} = c_{1,\uparrow}^\dag c_{2,\downarrow}^\dag c_{3,\uparrow}^\dag \ket{\rm Vac}$. In the quantum-circuit Monte Carlo (first-order leading-order-rotation formula), we take $\Delta t = 0.05$ and $N_{\rm s} = 10000$. In the classical Monte Carlo (zeroth-order Pauli-operator-expansion formula), we take $\Delta t = 0.01$ and $N_{\rm s} = 100000$~\cite{footnote}. The samples are generated according to the distribution in Eq.~(\ref{eq:pro}).
}
\label{fig:Hubbard}
\end{center}
\end{figure}

As shown in verified quantum phase estimation~\cite{OBrien2020} and dual-state purification~\cite{Huo2021}, a quantum circuit with the forward-backward structure incorporating postselection is robust to errors. For the postselection, we measure the $n$ qubits representing the system in addition to the ancillary qubit see Fig.~\ref{fig:circuit}(b). We only select the state when the measurement outcome is $\ket{0}^{\otimes n}$, which transforms the final state in Eq.~(\ref{eq:Phi}) into
\begin{eqnarray}
\ket{\Phi'} &=& \frac{\ket{0}_{\rm a} + \bra{\psi_{\rm f}} O_{\bfs} \ket{\psi_{\rm i}} \ket{1}_{\rm a}}{\sqrt{1+\absLR{\bra{\psi_{\rm f}} O_{\bfs} \ket{\psi_{\rm i}}}^2}} \otimes \ket{0}^{\otimes n}.
\label{eq:Phip}
\end{eqnarray}
Measuring the ancillary qubit in the state after postselection, we have
\begin{subequations}
\begin{eqnarray}
\mean{X_a}_{0} &=& \frac{2\Re \left( \bra{\psi_{\rm f}} O_{\bfs} \ket{\psi_{\rm i}} \right)}{1+\absLR{\bra{\psi_{\rm f}} O_{\bfs} \ket{\psi_{\rm i}}}^2},
\end{eqnarray}
\begin{eqnarray}
\mean{Y_a}_{0} &=& \frac{2\Im \left( \bra{\psi_{\rm f}} O_{\bfs} \ket{\psi_{\rm i}} \right)}{1+\absLR{\bra{\psi_{\rm f}} O_{\bfs} \ket{\psi_{\rm i}}}^2},
\end{eqnarray}
\begin{eqnarray}
\mean{Z_a}_{0} &=& \frac{1-\absLR{\bra{\psi_{\rm f}} O_{\bfs} \ket{\psi_{\rm i}}}^2}{1+\absLR{\bra{\psi_{\rm f}} O_{\bfs} \ket{\psi_{\rm i}}}^2},
\end{eqnarray}
\end{subequations}
where $\mean{\bullet}_{0}$ denotes the mean conditioned on the outcome $\ket{0}^{\otimes n}$. Solving the equations, we obtain
\begin{eqnarray}
\bra{\psi_{\rm f}} O_{\bfs} \ket{\psi_{\rm i}} = \frac{\mean{X_a}_{0} + i\mean{Y_a}_{0}}{1 + \mean{Z_a}_{0}}.
\label{eq:OssII}
\end{eqnarray}

The postselection forces most of qubits into a pure state, which eliminates errors that transform $\ket{0}^{\otimes n}$ into orthogonal states. In addition to postselection, we can purify the ancillary qubit as follows. According to Eq.~(\ref{eq:Phip}), the state of the ancillary qubit is a pure state when the quantum circuit is error-free. In the tomography purification, we implement the state tomography to the ancillary qubit and compute the eigenstate with the largest eigenvalue of the reconstructed reduced density matrix~\cite{Huo2021}. Using the eigenstate to compute the three means $\mean{\bullet}_{0}$, we can make sure that the final result is obtained from a pure state. In Sec.~\ref{sec:numeric}, we demonstrate that the inherent error mitigation can significantly reduce the error in QCMC.

Now, we have two protocols using the circuit in Fig.~\ref{fig:circuit}(b) to evaluate $\bra{\psi_{\rm f}} O_{\bfs} \ket{\psi_{\rm i}}$. In the protocol without postselection (see Sec.~\ref{sec:FBcircuit}), the estimator of $\bra{\psi_{\rm f}} O_{\bfs} \ket{\psi_{\rm i}}$ is unbiased, and it is optimal to take $M_{\rm s} = 1$. In the protocol with postselection, the estimator is biased due to the denominator in Eq.~(\ref{eq:OssII}) i.e.~the mean of estimates is not exactly $\bra{\psi_{\rm f}} O_{\bfs} \ket{\psi_{\rm i}}$ when $M_{\rm s}$ is finite. Therefore, for the postselection protocol, it is necessary to choose a large $M_{\rm s}$ to evaluate each $\mean{\bullet}_{0}$ (such that the bias is small) in order to obtain an accurate final result of the transition amplitude.

The inherent error mitigation increases the variance of QCMC. When the circuit is error-free, the postselection succeeds with the probability
\begin{eqnarray}
P_S = \frac{1}{2}\left(1+\absLR{\bra{\psi_{\rm f}} O_{\bfs} \ket{\psi_{\rm i}}}^2\right) \geq \frac{1}{2}.
\end{eqnarray}
If the circuit is implemented for $M_{\rm s}$ shots, only $P_S M_{\rm s}$ shots generate effective data on average. When the circuit is noisy, errors transform $\ket{0}^{\otimes n}$ into orthogonal states, which reduces the success rate. Therefore, the number of effective shots decreases with the error rate and the gate number, which causes an enlarged variance.

\subsubsection{Numerical demonstration}
\label{sec:numeric}

\begin{figure}[tbp]
\begin{center}
\includegraphics[width=1\linewidth]{\figpath/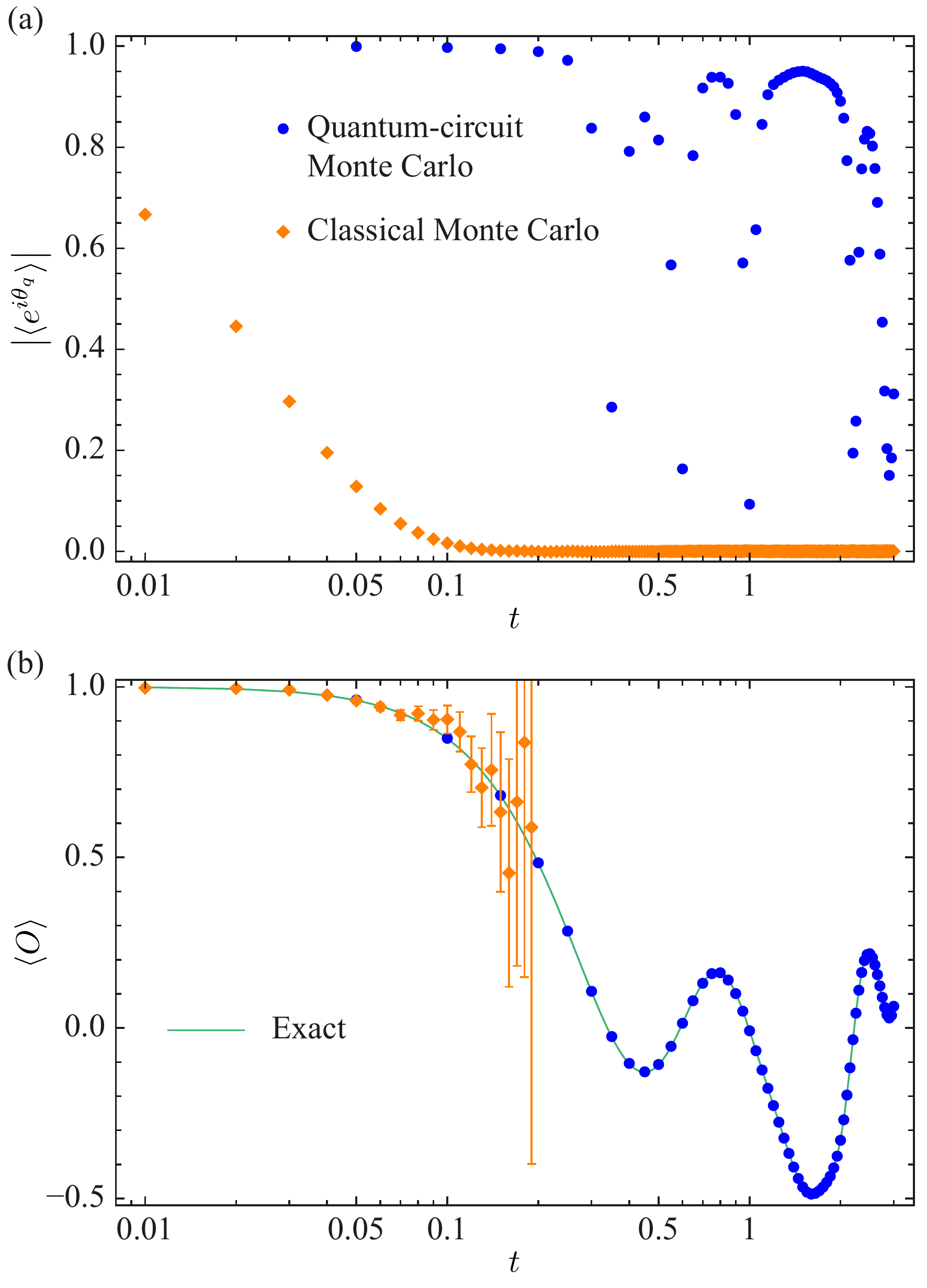}
\caption{
(a) The phase average and (b) the expected value of the observable $O = Z_3$ in the Monte Carlo simulation of the Heisenberg model. The Hamiltonian is $H = - J\sum_{i=1}^{N_S-1} (X_iX_{i+1}+Y_iY_{i+1}+Z_iZ_{i+1}) - h\sum_{i=1}^{N_S} Z_i$, where the number of spins is $N_S = 6$, and $J=h=1$. The simulation is to compute $\mean{O} = \bra{\psi_{\rm i}}e^{iHt}Oe^{-iHt}\ket{\psi_{\rm i}}$, where $\ket{\psi_{\rm i}} = \ket{010101}$. In the quantum-circuit Monte Carlo (first-order leading-order-rotation formula), we take $\Delta t = 0.05$ and $N_{\rm s} = 10000$. In the classical Monte Carlo (zeroth-order Pauli-operator-expansion formula), we take $\Delta t = 0.01$ and $N_{\rm s} = 200000$~\cite{footnote}. The samples are generated according to the distribution in Eq.~(\ref{eq:pro}).
}
\label{fig:Heisenberg}
\end{center}
\end{figure}

To demonstrate the inherent error mitigation, we consider the one-dimensional Fermi-Hubbard model and numerically simulate the noisy quantum computing on a classical computer. The Hamiltonian reads
\begin{eqnarray}
H &=& - J \sum_{i=1}^{N_L-1} \sum_{s=\uparrow,\downarrow} \left( c_{i,s}^\dag c_{i+1,s} + c_{i+1,s}^\dag c_{i,s} \right) \notag \\
&&+ U\sum_i c_{i,\uparrow}^\dag c_{i,\uparrow} c_{i,\downarrow}^\dag c_{i,\downarrow},
\label{eq:FHM}
\end{eqnarray}
where $N_L = 3$ is the number of sites and $c_{i,s}$ is the annihilation operator for the fermion with spin-$s$ on the $i$th site. This model can be encoded into $2N_L$ qubits using the Jordan-Wigner transformation.

We use the first-order Lie-Trotter-Suzuki product formula and a corresponding summation formula to simulate the real time evolution. The initial state is $\ket{\psi_{\rm i}} = c_{1,\uparrow}^\dag c_{2,\downarrow}^\dag c_{3,\uparrow}^\dag \ket{\rm Vac}$, where $\ket{\rm Vac}$ is the vacuum state, and we take $\ket{\psi_{\rm f}} = \ket{\psi_{\rm i}}$. The observable is $O = 2c_{3,\uparrow}^\dag c_{3,\uparrow} - \openone$ and the simulation is to compute $\mean{O} = \bra{\psi_{\rm i}}e^{iHt}Oe^{-iHt}\ket{\psi_{\rm i}}$. To minimise the variance of QCMC, we first expand the correction operator using Pauli operators, i.e.
\begin{eqnarray}
V_1 = \sum_{\sigma\in\bfP_n} (a_\sigma - ib_\sigma) \sigma,
\end{eqnarray}
where $a_\sigma$ and $b_\sigma$ are real, and
\begin{eqnarray}
a_\sigma - ib_\sigma = 2^{-n}\Tr\left[\sigma e^{-iH\Delta t}S_1(\Delta t)^\dag\right].
\end{eqnarray}
We have $a_{\openone} > 0$ and $b_{\openone} = 0$. Then, we take the summation formula
\begin{eqnarray}
e^{-iH\Delta t} = \sum_{\sigma\in \bfP_n\setminus\{\openone\}} \left( a_\sigma \sigma + \beta_\sigma e^{-i{\rm sgn}(b_\sigma)\phi \sigma} \right) S_1,
\end{eqnarray}
where $\phi = \arctan(a_{\openone}^{-1}\sum_\sigma \abs{b_\sigma})$ and $\beta_\sigma = \abs{b_\sigma}/\sin\phi$.

Using the forward-backward circuit for error mitigation, we find that the impact of machine errors can be significantly suppressed, as shown in Fig.~\ref{fig:FHM}. We model the noise in quantum computing using the depolarising error model. For a controlled-NOT gate, the noise map is given by Eq.~(\ref{eq:noise}) with parameters $p_\sigma = p/15$. We neglect errors in the initialisation, single-qubit gates, and measurement. In the numerical simulation, we take the error rate per gate $p = 0.03\%$. The number of controlled-NOT gates for each $S_1$ is $14$ and the simulation involves at most $85$ time steps, i.e.~the total number of controlled-NOT gates is above $2380$. Therefore, the maximum total error rate is above $71.4\%$. After the error mitigation, we find that the overall accuracy of the summation formula taking $N_{\rm s} = 10000$ samples is higher than the product formula without machine errors.


\section{Quantum computing versus classical computing}
\label{sec:QandC}

Sampling noise is the main source of error in the QCMC algorithm. The Monte Carlo variance increases exponentially with the evolution time as approximately $\frac{1}{N_{\rm s}} C_A^{4N} = \frac{1}{N_{\rm s}} e^{4t \Delta t^{-1}\ln C_A}$. As summarised in Table.~\ref{table:CA}, $C_A = 1+\xi \Delta t^k + O(\Delta t^{k+1})$, where $\xi$ is a constant depending on the Hamiltonian. When $k>1$, by taking $\Delta t = \left(\frac{\delta}{4t\xi}\right)^{1/(k-1)}$, we can reduce the factor to $C_A^{4N} = e^{\delta + O\left(\delta^{k/(k-1)}\right)}$ for any small $\delta$. We note that $k>1$ in $l$th-order POE formulas with $l>0$ and all LOR formulas. For the zeroth-order POE formula, because $C_A = e^{h_{\rm tot}\Delta t}$ (i.e.~$k=1$), we have $C_A^{4N} = e^{4h_{\rm tot}t}$ for all $\Delta t$.

It is widely believed that a classical computer cannot simulate the time evolution of general quantum many-body systems at a polynomial cost, which is one of main motivations for quantum computing~\cite{Feynman1982}. The QCMC algorithm with the zeroth-order POE formula is equivalent to a classical algorithm, i.e.~Green's function Monte Carlo taking the computational basis, for a large class of Hamiltonians (see Sec.~\ref{sec:CC}). In this classical algorithm, the variance increases exponentially with the evolution time and system size, i.e.~approximately $\frac{1}{N_{\rm s}} e^{4 h_{\rm tot}t}$, whatever $\Delta t$ we choose. Here, $t$ is the evolution time and $h_{\rm tot}$ increases with the system size. The variance is up to minimization, e.g.~changing the Hilbert space basis~\cite{Hangleiter2020} and optimising the method for generating samples. Nevertheless, the existence of a generic approach that reduces the exponential scaling to polynomial one is unlikely~\cite{Troyer2005}.

On a fault-tolerant quantum computer, we can simulate the time evolution of quantum many-body systems at a polynomial cost. By taking a small $\Delta t$, we can reduce the factor $C_A^{4N}$ to a satisfactory level and $\Delta t$ scales polynomially with $t$ and $h_{\rm tot}$. Therefore, the number of times steps $N = t/\Delta t$, i.e.~the circuit depth, scales polynomially with $t$ and $h_{\rm tot}$.

To demonstrate the impact on the sign problem in QMC incorporating quantum computing, we simulate the real time evolution of two models, the Fermi-Hubbard model and the Heisenberg model. We use two formulas in the simulation of each model, the zeroth-order POE and first-order LOR formulas. The zeroth-order POE formula corresponds to a classical QMC algorithm. Because the zeroth-order POE formula only includes products of Pauli operators, we can efficiently evaluate it on a classical computer even when the system size is large. The first-order LOR formula includes products of non-Pauli unitary operators, e.g.~the product in Eq.~(\ref{eq:S1}). With a quantum computer, we can efficiently evaluate these non-Pauli products when the system is large. For the purpose of comparing the sign problem in two formulas, we evaluate both formulas on a classical computer given that the system size is up to six qubits. The phase average $\langle e^{i\theta_q} \rangle$ is used to indicate the sign problem, where $e^{i\theta_q}$ denotes the phase of $e^{i\theta_{s}}\bra{\psi_{\rm f}} O_{\bfs} \ket{\psi_{\rm i}}$. We find that the sign problem is significant in the zeroth-order POE formula, i.e.~$\langle e^{i\theta_q} \rangle$ converges to zero rapidly with the evolution time (see Figs.~\ref{fig:Hubbard} and \ref{fig:Heisenberg}). As a result, the estimation of an observable has a large variance. In comparison, the sign problem is mild in the first-order LOR formula, i.e.~$\langle e^{i\theta_q} \rangle$ is finite. With the sign problem mitigated, the simulation using the first-order LOR formula is accurate for a much longer evolution time compared with the zeroth-order POE formula.

\begin{figure}[tbp]
\begin{center}
\includegraphics[width=1\linewidth]{\figpath/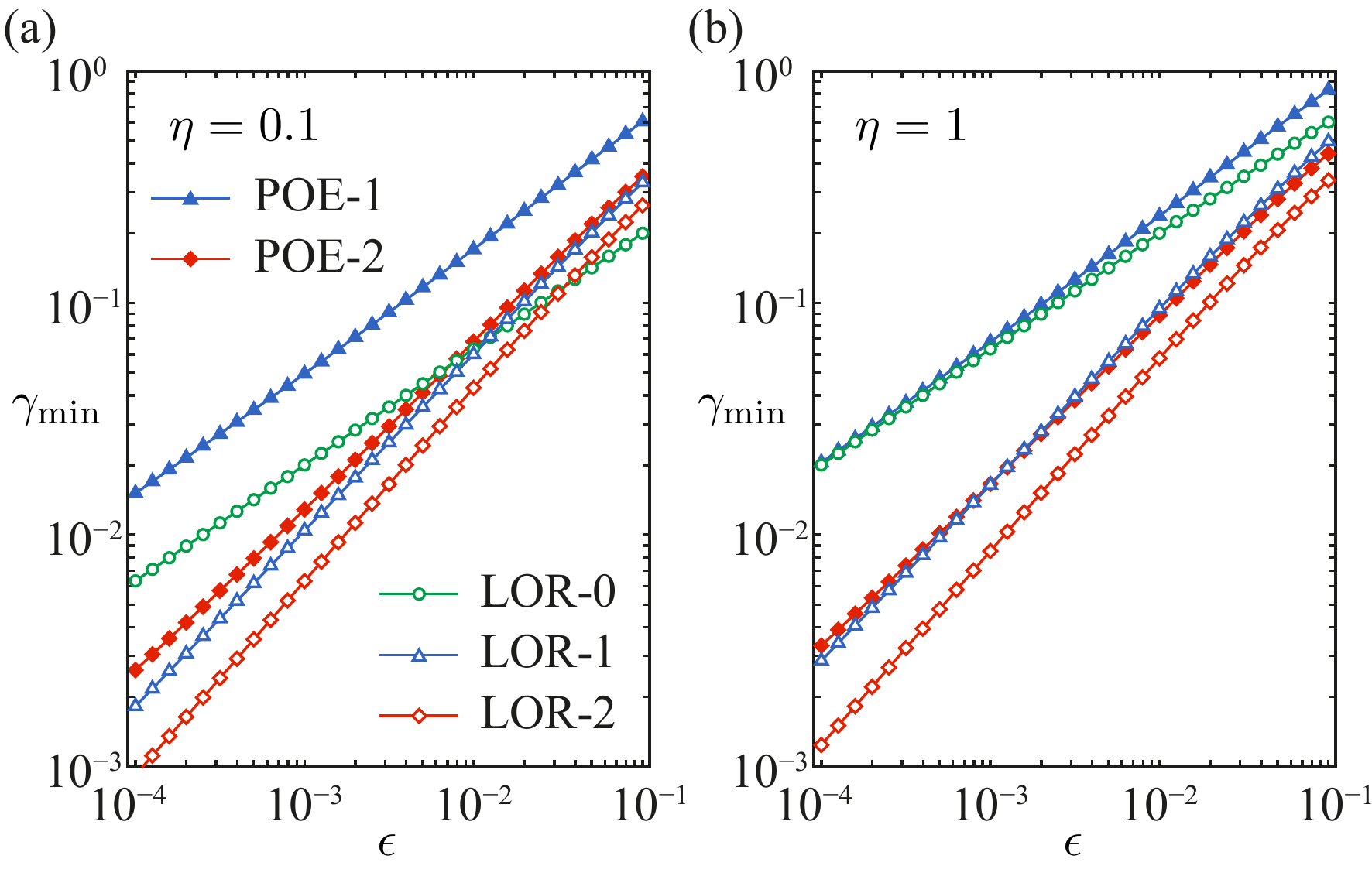}
\caption{
The minimum rate of increase of variance $\gamma_{\rm min}$. $\epsilon$ is the error rate of one elementary Lie-Trotter-Suzuki product $S_1$. $\eta \epsilon$ is the average error rate of the controlled-correction gates in one time step. POE $l$ denotes the $l$th-order Pauli-operator-expansion formula and LOR $l$ denotes the $l$th-order leading-order-rotation formula.
}
\label{fig:scaling}
\end{center}
\end{figure}

On a noisy quantum computer, the cost of simulating quantum many-body systems increases exponentially with the evolution time and system size and the rate of increase decreases with the error rate. Using the quasi-probability decomposition to mitigate errors, the error mitigation enlarges the variance. The variance taking into account quantum error mitigation is approximately $\frac{1}{N_{\rm s}} C_A^{4N} C_E^2$. We consider the compact circuit in Fig.~\ref{fig:circuit}(a). Let $\epsilon$ be the error rate of one elementary product $S_1$ (we assume that $S'_1$ has the same error rate), let $g$ be the number of $S_1$ and $S_1'$ products per time step, and let $\eta \epsilon$ be the average error rate of the controlled-correction gates in one time step. The total error rate of one time step is approximately $(g+\eta)\epsilon$. Here, $g=0,1,2$ for zeroth-, first-, and second-order formulas, respectively. Suppose that the total error rate of other operations [which are out of the bracket in Fig.~\ref{fig:circuit}(a)] is $\epsilon'$ and the factor due to error mitigation is $C_E \simeq (1+2\epsilon')\left[1+2(g+\eta)\epsilon\right]^N$, according to the Pauli error model. Then, we can express the variance in the form
\begin{eqnarray}
\frac{1}{N_{\rm s}} C_A^{4N} C_E^2 \simeq \frac{1}{N_{\rm s}} (1+2\epsilon')^2 e^{4\gamma h_{\rm tot}t},
\end{eqnarray}
where
\begin{eqnarray}
\gamma  = \frac{1}{h_{\rm tot}\Delta t} \ln \left[C_A\sqrt{1+2(g+\eta)\epsilon}\right].
\end{eqnarray}
We find that the rate $\gamma$ decreases with the error rate.

Given the error rate of the noisy quantum computer, we choose the time step size $\Delta t$ to minimise the rate $\gamma$. Taking $C_A \simeq 1+\xi\Delta t^{k}$, we find that the optimal step size is
\begin{eqnarray}
\Delta t_{\rm opt} \simeq \left[\frac{(g+\eta)\epsilon}{(k-1)\xi}\right]^{1/k},
\end{eqnarray}
and the corresponding minimum rate is
\begin{eqnarray}
\gamma_{\rm min} \simeq \frac{k\xi^{1/k}}{h_{\rm tot}} \left[\frac{(g+\eta)\epsilon}{k-1}\right]^{\frac{k-1}{k}}.
\end{eqnarray}
For the second-order LOR formula, $k = 6$, $g = 2$, and $\xi < h_{\rm tot}^6/(2\times 18^2)$. In Fig.~\ref{fig:scaling}, we plot the minimum rate computed numerically using formulas of $C_A$ in Table.~\ref{table:CA}. For the first- and second-order formulas, we take the upper bound of the simplified leading-order contribution. We find that the second-order LOR formula outperforms other formulas. Taking higher-order formulas does not further reduce $\gamma$ for the given error rates.

Although the cost of the QCMC algorithm on a noisy quantum computer scales exponentially with the evolution time and system size in the same way as the classical algorithm, the quantum computing can accelerate QMC by reducing the variance. To achieve the computation accuracy $\delta$, i.e.~to reduce the variance to $\delta^2$, we take $N_{\rm s} \sim e^{4\gamma h_{\rm tot}t}/\delta^2$. In the classical algorithm, $\gamma = 1$. In the quantum algorithm, taking the minimum value for $\eta = 1$ [see Fig.~\ref{fig:scaling}(b)], we have $\gamma \simeq 0.34$ when the error rate per elementary product is $\epsilon = 0.1$ and $\gamma \simeq 0.058$ when $\epsilon = 0.01$. For $h_{\rm tot}t = 4$, the quantum algorithm reduces the sample size $N_{\rm s}$ by a factor of approximately $4\times 10^4$ when $\epsilon = 0.1$ and approximately $4\times 10^6$ when $\epsilon = 0.01$. The advantage of the quantum algorithm grows when the error rate decreases: fitting to the second-order LOR curve in Fig.~\ref{fig:scaling}(b), we have $\gamma \simeq 2.45 \epsilon^{0.82}$.

We can optimise the quantum algorithm to reduce the variance in various ways. First, we can significantly reduce $C_A$ for a Hamiltonian with only local interactions. $C_A$ is greater than one because of the correction operator, which is used to compensate the difference between the exact time evolution operator and the Lie-Trotter-Suzuki product. According to the Baker-Campbell-Hausdorff formula, this difference is a series of commutators. For local interactions, most of the commutators in low-order terms are zero. In this case, expanding the correction operator according to the Baker-Campbell-Hausdorff formula (instead of the direct Taylor expansion) can reduce $C_A$. Second, similar to the classical algorithm, with some knowledge of $\bra{\psi_{\rm f}} O_{\bfs} \ket{\psi_{\rm i}}$, we can optimise the distribution of generating samples to reduce the variance. Third, the variance due to quantum error mitigation can be reduced. In the error mitigation protocol used to estimate $\gamma$, we correct all Pauli errors in the circuit, which is unnecessary. Because the ancillary qubit is measured to evaluate the transition amplitude, we only need to correct errors that affect the ancillary qubit. These errors can be identified and corrected by utilising the learning-based approach of error mitigation~\cite{Strikis2020}.

\section{Conclusions}
\label{sec:conclusions}

In this paper, we propose a QMC algorithm that uses quantum computing as a subroutine, which allows the non-variational quantum simulation to be implemented with noisy intermediate-scale quantum hardware. In our algorithm, we use exact summation formulas to express the time evolution operator. We optimise these summation formulas and quantum circuits to minimise the Monte Carlo variance and circuit depth. The optimal distribution of generating samples in Monte Carlo is derived in the circumstances of probabilistic evaluation using quantum computing. On a noisy quantum computer, we can use probabilistic error cancellation or inherent error mitigation to eliminate machine errors. By choosing the parameter $\Delta t$, we can maximise the quantum speedup given a finite error rate. This scheme illustrates a way of designing quantum algorithms with reduced circuit depth by using Monte Carlo techniques~\cite{Casares2021}.

Our algorithm shows that a quantum computer without fault tolerance can speed up solving practical problems. In terms of algorithmic complexity, quantum computing has an advantage over classical computing in many computational tasks, in the fault-tolerance regime achieved with quantum error correction~\cite{Bravyi2020}. Even without error correction, a quantum device can perform tasks that are intractable for classical computers, such as sampling the output of a quantum circuit~\cite{Arute2019}. Our algorithm is to solve a practical problem, i.e.~the non-variational simulation of quantum many-body systems. We theoretically analyse the complexity of our algorithm, i.e. the circuit depth and sampling cost. The complexity is polynomial on a fault-tolerant quantum computer. On a noisy quantum computer, although the complexity is exponential due to the finite error rate, our algorithm can still outperform classical algorithms and speed up Monte Carlo calculations by substantially reducing the sign problem.

\begin{acknowledgments}
We acknowledge the use of simulation toolkit QuESTlink~\cite{Jones2020} for this work.
We acknowledge the support of the National Natural Science Foundation of China (Grant No. 11875050 and 12088101) and NSAF (Grant No. U1930403).
\end{acknowledgments}

{\it Note added.}---Shortly after this work (the first version) posted on arXiv, a relevant paper was also made public~\cite{Huggins2021}, which reports a quantum algorithm for imaginary time dynamics based on QMC.

\appendix

\section{Leading-order terms}
\label{app:LOT}

Let $A_j \equiv -i h_j \sigma_j \Delta t$ and $A \equiv \sum_i A_i = -iH\Delta t$ for simplicity. The Taylor expansion of the time evolution operator reads
\begin{eqnarray}
e^{-iH\Delta t} &=& e^A = \openone + A + \frac{1}{2} A^2 + \frac{1}{6} A^3 + O\left(\Delta t^4\right).~~
\end{eqnarray}
We have
\begin{eqnarray}
A^2 = \sum_{i<j} \left(A_i A_j + A_j A_i\right) + \sum_{i} A_i^2
\end{eqnarray}
and
\begin{eqnarray}
A^3 &=& \sum_{i<j<k} \left( A_i A_j A_k + A_k A_j A_i + A_j A_i A_k \right. \notag \\
&&\left. + A_k A_i A_j + A_i A_k A_j + A_j A_k A_i \right) \notag \\
&&+ \sum_{i<j} \left( A_i^2 A_j + A_j A_i^2 + A_i A_j A_i \right. \notag \\
&&\left. + A_i A_j^2 + A_j^2 A_i + A_j A_i A_j \right) \notag \\
&&+ \sum_{i} A_i^3.
\end{eqnarray}

\subsection{First-order formula}

According to the first-order formula, we have
\begin{eqnarray}
S_{1}(\Delta t)^\dag &=& e^{-A_1} \cdots e^{-A_M} \notag \\
&=& \prod_{i=1}^M \left( \openone - A_i + \frac{1}{2} A_i^2 - \frac{1}{6} A_i^3 + O\left(\Delta t^4\right) \right) \notag \\
&=& \openone - A + \frac{1}{2} A^{(2)} - \frac{1}{6} A^{(3)} + O\left(\Delta t^4\right),
\end{eqnarray}
where
\begin{eqnarray}
A^{(2)} &=& 2\sum_{i<j} A_i A_j + \sum_{i} A_i^2, \\
A^{(3)} &=& 6\sum_{i<j<k} A_i A_j A_k + 3\sum_{i<j} \left( A_i^2 A_j + A_i A_j^2 \right) \notag \\
&&+ \sum_{i} A_i^3.
\end{eqnarray}
The correction operator is
\begin{eqnarray}
V_{1}(\Delta t) &=& \openone - A^2 + \frac{1}{2}\left(A^2+A^{(2)}\right) \notag \\
&&+ \frac{1}{2}A\left(A^{(2)} - A^2\right) + \frac{1}{6}\left(A^3-A^{(3)}\right) \notag \\
&&+ O\left(\Delta t^4\right).
\end{eqnarray}
Then, we have
\begin{eqnarray}
F_{1}^{(2)}(\Delta t) &=& - A^2 + \frac{1}{2}\left(A^2+A^{(2)}\right) = \frac{1}{2}\left(A^{(2)} - A^2\right) \notag \\
&=& \frac{1}{2}\sum_{i<j} \left(A_i A_j - A_j A_i\right)
\label{eq:F1}
\end{eqnarray}
and
\begin{eqnarray}
F_{1}^{(3)}(\Delta t) &=& \frac{1}{2}A\left(A^{(2)} - A^2\right) + \frac{1}{6}\left(A^3-A^{(3)}\right) \notag \\
&=& \frac{1}{2} \sum_{i<j<k} \left( A_i A_j A_k - A_k A_j A_i + A_j A_i A_k \right. \notag \\
&&\left. + A_k A_i A_j - A_i A_k A_j - A_j A_k A_i\right) \notag \\
&&+ \frac{1}{2} \sum_{i<j} \left( A_i^2 A_j - A_i A_j A_i - A_j^2 A_i + A_j A_i A_j \right) \notag \\
&&+ \frac{1}{6}\left(A^3-A^{(3)}\right) \notag \\
&=& \frac{1}{6} \sum_{i<j<k} \left( - 2A_i A_j A_k - 2A_k A_j A_i + 4A_j A_i A_k \right. \notag \\
&&\left. + 4A_k A_i A_j - 2A_i A_k A_j - 2A_j A_k A_i\right) \notag \\
&&+ \frac{1}{6} \sum_{i<j} \left( A_i^2 A_j + A_j A_i^2 - 2A_i A_j A_i \right. \notag \\
&&\left. - 2A_i A_j^2 - 2A_j^2 A_i + 4A_j A_i A_j \right).
\end{eqnarray}
According to Eq.~(\ref{eq:F1}), the contribution of $F_{1}^{(2)}(\Delta t)$ to the normalisation factor is
\begin{eqnarray}
\sum_{i<j} \abs{h_i h_j} \Delta t^2 < \frac{1}{2} \left(\sum_{i} \abs{h_i}\right)^2 \Delta t^2.
\end{eqnarray}

\subsection{Second-order formula}

We can write the second-order correction operator as
\begin{eqnarray}
V_{2}(\Delta t) &=& V_{1}\left(-\frac{\Delta t}{2}\right)^\dag V_{1}\left(\frac{\Delta t}{2}\right) \notag \\
&=& \openone + F_{1}^{(3)}\left(\frac{\Delta t}{2}\right) + F_{1}^{(3)}\left(-\frac{\Delta t}{2}\right)^\dag \notag \\
&&+ O(\Delta t^5).
\end{eqnarray}
Then, we have
\begin{eqnarray}
F_{2}^{(3)}(\Delta t) &=& F_{1}^{(3)}\left(\frac{\Delta t}{2}\right) + F_{1}^{(3)}\left(-\frac{\Delta t}{2}\right)^\dag.
\end{eqnarray}
Accordingly, the contribution of $F_{2}^{(3)}(\Delta t)$ to the normalisation factor is
\begin{eqnarray}
&&\frac{1}{3} \sum_{i<j<k} \abs{h_i h_j h_k} \Delta t^3 + \frac{1}{12} \sum_{i<j} \left(\abs{h_i^2 h_j} + 2\abs{h_i h_j^2}\right)\Delta t^3 \notag \\
&<& \frac{1}{18} \left(\sum_{i} \abs{h_i}\right)^3 \Delta t^3.
\end{eqnarray}

\section{Fermi-Hubbard model}
\label{app:FHM}

The Hamiltonian of Fermi-Hubbard model reads
\begin{eqnarray}
H_{\rm FH} &=& - \sum_{i<j} J_{i,j} \sum_{s=\uparrow,\downarrow} \left( c_{i,s}^\dag c_{j,s} + c_{j,s}^\dag c_{i,s} \right) \notag \\
&&+ U\sum_i \left(c_{i,\uparrow}^\dag c_{i,\uparrow} - \frac{\openone}{2}\right)\left(c_{i,\downarrow}^\dag c_{i,\downarrow} - \frac{\openone}{2}\right),
\end{eqnarray}
where $c_{i,s}$ is the annihilation operator for the fermion with spin-$s$ on the $i$th site. Operators of fermions satisfy $\{c_{i,s},c_{i',s'}\} = 0$ and $\{c_{i,s},c_{i',s'}^\dag\} = \delta_{i,i'}\delta_{s,s'}\openone$. Here, we modify the original Fermi-Hubbard model by adding a uniform on-site potential $-\frac{U}{2} N$, which does not affect the time evolution if the initial state is an eigenstate of the total particle number operator $N = \sum_{i,s} c_{i,s}^\dag c_{i,s}$. For a bipartite lattice, $J_{i,j} = 0$ for all $i+j\in \mathrm{Even}$, i.e.~two sites are not coupled if their labels have the same parity.

To encode the Fermi-Hubbard model into qubits, we take the Jordan-Wigner transformation
\begin{eqnarray}
c_{i,\uparrow} &=& \frac{Y_{2i-1}-iZ_{2i-1}}{2} \prod_{l<2i-1} X_l, \notag \\
c_{i,\downarrow} &=& \frac{Y_{2i}-iZ_{2i}}{2} \prod_{l<2i} X_l,
\end{eqnarray}
where $X_a$, $Y_a$, and $Z_a$ are the Pauli operators of qubit $a$. The spin-$\uparrow$ and the spin-$\downarrow$ on the $i$th site are encoded on the qubits $(2i-1)$ and $2i$, respectively. According to the Jordan-Wigner transformation, the qubit Hamiltonian of Fermi-Hubbard model is
\begin{eqnarray}
H_{\rm FH} &=& - \sum_{i<j} \frac{J_{i,j}}{2} \left(Y_{2i-1,2j-1} + Z_{2i-1,2j-1}\right. \notag \\
&&+ \left.Y_{2i,2j} + Z_{2i,2j}\right) + \frac{U}{4} \sum_i X_{2i-1}X_{2i},
\label{eq:HFH}
\end{eqnarray}
where
\begin{eqnarray}
Y_{a,b} &=& Y_aY_b \prod_{a<l<b}X_l, \notag \\
Z_{a,b} &=& Z_aZ_b \prod_{a<l<b}X_l.
\end{eqnarray}

Each Pauli operator $\sigma$ corresponds to an $x$ binary string according to Eq.~(\ref{eq:sigma}), and we define $\bfx(\sigma) \equiv (x_1,\ldots,x_n)$ as the $x$ binary string of the Pauli operator $\sigma$. A Hamiltonian does not have short-time interference if $\bfx(\sigma_1)\neq \bfx(\sigma_2)$ for any pair of Pauli-operator terms $\sigma_1$ and $\sigma_2$ in the Hamiltonian. We use $\bfx_{a,b}$ to denote the binary string for which $x_k = 0$ if $k<a$ or $k>b$ and $x_k = 1$ if $a\leq k\leq b$. Then,
\begin{eqnarray}
\bfx\left(Y_{a,b}\right) &=& \bfx_{a,b}, \notag \\
\bfx\left(Z_{a,b}\right) &=& \bfx_{a+1,b-1}, \notag \\
\bfx\left(X_{2i-1}X_{2i}\right) &=& \bfx_{2i-1,2i}.
\end{eqnarray}

We find that $x$ strings of $Y_{a,b}$ and $Z_{a,b}$ terms in Eq.~(\ref{eq:HFH}) are all different from $X_{2i-1}X_{2i}$ terms: note that $a$ and $b$ have the same parity. The only question is whether $Y_{a,b}$ and $Z_{a',b'}$ have the same $x$ string. If their $x$ strings are the same, we must have $a = a'+1$ and $b = b'-1$. For a bipartite lattice, $\frac{b-a}{2}$ and $\frac{b'-a'}{2}$ are both odd: however, $\frac{b'-1-a'-1}{2}$ is even if $\frac{b'-a'}{2}$ is odd. Therefore, $x$ strings of $Y_{a,b}$ and $Z_{a',b'}$ are always different.


\begin{thebibliography}{9}
\bibitem{footnote} The variance of the zeroth-order POE formula is independent of $\Delta t$. Amplitudes $\bra{\psi_{\rm f}} O_{\bfs} \ket{\psi_{\rm i}}$ in the classical Monte Carlo simulation take three values $0$ and $\pm 1$. We choose sample numbers $N_{\rm s}$ in the classical Monte Carlo such that the number of non-zero amplitudes is larger than $10000$. In other words, even if we generate $10000$ samples according to the optimal distribution in Eq.~(\ref{eq:optimalPRO}), the variance is not smaller.
\end{thebibliography}
\end{document}